\documentclass{aa}  
\usepackage{graphicx}
\usepackage{adjustbox}
\usepackage{multirow}
\usepackage{xcolor}
\usepackage{hyperref}
\hypersetup{
    colorlinks = true,
    citecolor = {blue},
    linkcolor = {red},
    urlcolor = {blue}
}
\usepackage{media9}
\newcommand{\ii}{ {\rm i} }
\newcommand{\bxi}{\boldsymbol{\xi}}
\def\br{\mbox{\boldmath$r$}}
\def\bu{\mbox{\boldmath$u$}}
\newcommand{\vect}[1]{\boldsymbol{#1}}

\def\dd{{\rm d}}

\usepackage{natbib}
\bibpunct{(}{)}{;}{a}{}{,} % to follow the A&A style

\usepackage{txfonts}

\makeatletter
\renewcommand*\aa@pageof{, page \thepage{} of \pageref*{LastPage}}
\makeatother

\begin{document}

   \title{Contribution of flows around active regions to the north-south helioseismic travel-time measurements}
    \titlerunning{Active-region flows in north-south helioseismic measurements}

   \author{P.-L. Poulier \inst{1}
          \and
        Z.-C. Liang \inst{1}
        \and
        D. Fournier \inst{1}
        \and
        L. Gizon \inst{1,2}
          }

   \institute{Max-Planck-Institut f{\"u}r Sonnensystemforschung, Justus-von-Liebig-Weg 3, 37077 G{\"o}ttingen,
    Germany \\
    \email{poulier@mps.mpg.de}
              \and
              Georg-August-Universit{\"a}t, Friedrich-Hund-Platz 1, 37077 G{\"o}ttingen, Germany\\
             }

   \date{Received ; }

% \abstract{}{}{}{}{} 
% 5 {} token are mandatory
 
  \abstract
   {In local helioseismology, the travel times of acoustic waves propagating in opposite directions along the same meridian inform us about horizontal flows in the north-south direction. The longitudinal averages of the north-south helioseismic travel-time shifts vary with the sunspot cycle. }
   {We aim to study the contribution of inflows into solar active regions to this solar-cycle  variation. }
   {
   To do so, we identify the local flows around active regions in the horizontal flow maps obtained from correlation tracking of granulation in SDO/HMI continuum images.
   We compute the forward-modeled travel-time perturbations caused by these inflows using 3D sensitivity kernels. In order to compare with the observations, we average these forward-modeled travel-time perturbations over longitude and time in the same way as the measured travel times.
   }
   {
   The forward-modeling approach shows that the inflows
   associated with active regions may account for only a fraction of the solar-cycle variations in the north-south travel-time measurements.
  }
   {The travel-time perturbations caused by the large-scale inflows surrounding the active regions do not explain in full the solar-cycle variations seen in the helioseismic measurements of the meridional circulation.
   }


   \maketitle
%
%________________________________________________________________

%\tableofcontents

\section{Introduction \label{sec:intro}}

The Sun's meridional flow at the surface is poleward with a maximum  amplitude of about $15$ m/s \citep{1979Duvall}. The meridional ciculation, both at the surface and in the deep convection zone, is believed to be a key ingredient in flux-transport dynamo models \citep[e.g.,][]{1991Wang,2006Dikpati,2009Dikpati}.  Observationally, the meridional circulation is given as the longitudinal average of the north-south flows. This longitudinal average is not constant in time: its amplitude and latitudinal dependence change over the solar cycle \citep[e.g.,][]{1993Komm,2010Hathaway,2011Hathaway}. It has been proposed that extended inflows around solar active regions \citep{2001Gizon} modulate the meridional flow at the surface \citep{2004Gizon, 2010Gizon}. These inflows have amplitudes of up to $50$~m/s near the surface and extend up to $10^\circ$ from the center of the active regions or further \citep{2001Gizon,2003Hindman,2004Haber,2009Hindman,2019Braun,2021Gottschling}.

Helioseismic travel-time shifts in the north-south direction are sensitive to the meridional flow \citep[e.g.,][]{1997Giles}. The measurements made by \citet{2020Gizon} show a solar-cycle modulation. We aim to determine how much of this modulation may indeed be attributed to the near-surface active-region flows. To this end, we isolate and measure the active-region flows in local correlation tracking (LCT) flow maps.
By assuming a depth dependence of these flows, we compute forward helioseismic travel-time perturbations in the north-south direction to estimate their contribution to the fluctuations seen in the time-distance measurements. Section~\ref{sec:method} presents the data and the method. Section~\ref{sec:flows} shows the latitudinal and longitudinal components of the resulting active-region flows. We compute in Section~\ref{sec:traveltimes} the north-south forward helioseismic travel-time perturbations associated with these flows, using 3D Born sensitivity kernels. We attempt to model the inflows with a simple model based on the latitudinal gradient of the unsigned magnetic field in Section~\ref{sec:extension}, in order to extend the analysis to the previous solar cycle. We compare our results with helioseismic measurements in Section~\ref{sec:comparison}.

%__________________________________________________________________

\section{Horizontal flows from granulation tracking \label{sec:method}}
\subsection{Flow maps from LCT \label{sec:lct}}

We use the horizontal flow maps computed by \citet{2017Loeptien}. The original data set covered the period from May 2010 to April 2016, and has later been extended till April 2019 \citep{2021Gottschling}. The maps were obtained by using the Fourier Local Correlation Tracking code \citep{2004Welsch,2008Fisher}. The code tracked pairs of consecutive full-resolution intensity images from SDO/HMI \citep{2012Schou}, so that the flows represent the surface motions of solar granulation. The cadence of the flow maps is 30 minutes.

The data contained systematics like the orbital motions of SDO and the shrinking-Sun effect \citep{2004Lisle,2016Loeptien}. \citet{2017Loeptien} decomposed the flow maps into Zernike polynomials and filtered the time series of the coefficient amplitudes to remove the mean of the time series, the periods of 24 h, 1 year, and their corresponding harmonics, which takes care of most of the systematics.

The resulting data product contains the time-varying part of the rotation (torsional oscillations) and of the meridional circulation, plus potentially residual systematics \citep{2021Gottschling}. The velocities are in CCD-frame units (pixels per second). The size of these filtered images is $1024\times1024$ pixels; we perform a $5\times5$ binning on these images, yielding a spatial resolution of about 10 arcsec per pixel which corresponds to 0.6 heliographic degree per pixel at disk center.

\subsection{Construction of background in CCD frame}

\citet{2021Gottschling} reported that the LCT flow maps contain residual systematics. This means that large-scale background flows shall be removed from the flow maps in order to measure the flows related to active regions. These systematics likely depend on the position on the visible disk, in particular if they are a residual of the shrinking-Sun effect.
Therefore, the background
will be measured in the CCD frame; the procedure is described below.

\subsubsection{Contours around active-region flows \label{sec:contour_def}}
We define the background as the area located far away from the magnetic activity and its related flows.
In order to measure the background flows, we need to identify the active regions first.
We use the HMI magnetograms at the same time steps as the LCT data. We assume that the magnetic field is purely radial and thus divide the line-of-sight magnetic field $B_\mathrm{LOS}$ by the cosine of the great-circle angular distance to disk center to obtain $B_r$.

\begin{figure}[t]
    \centering
    \includegraphics[width=0.5\textwidth]{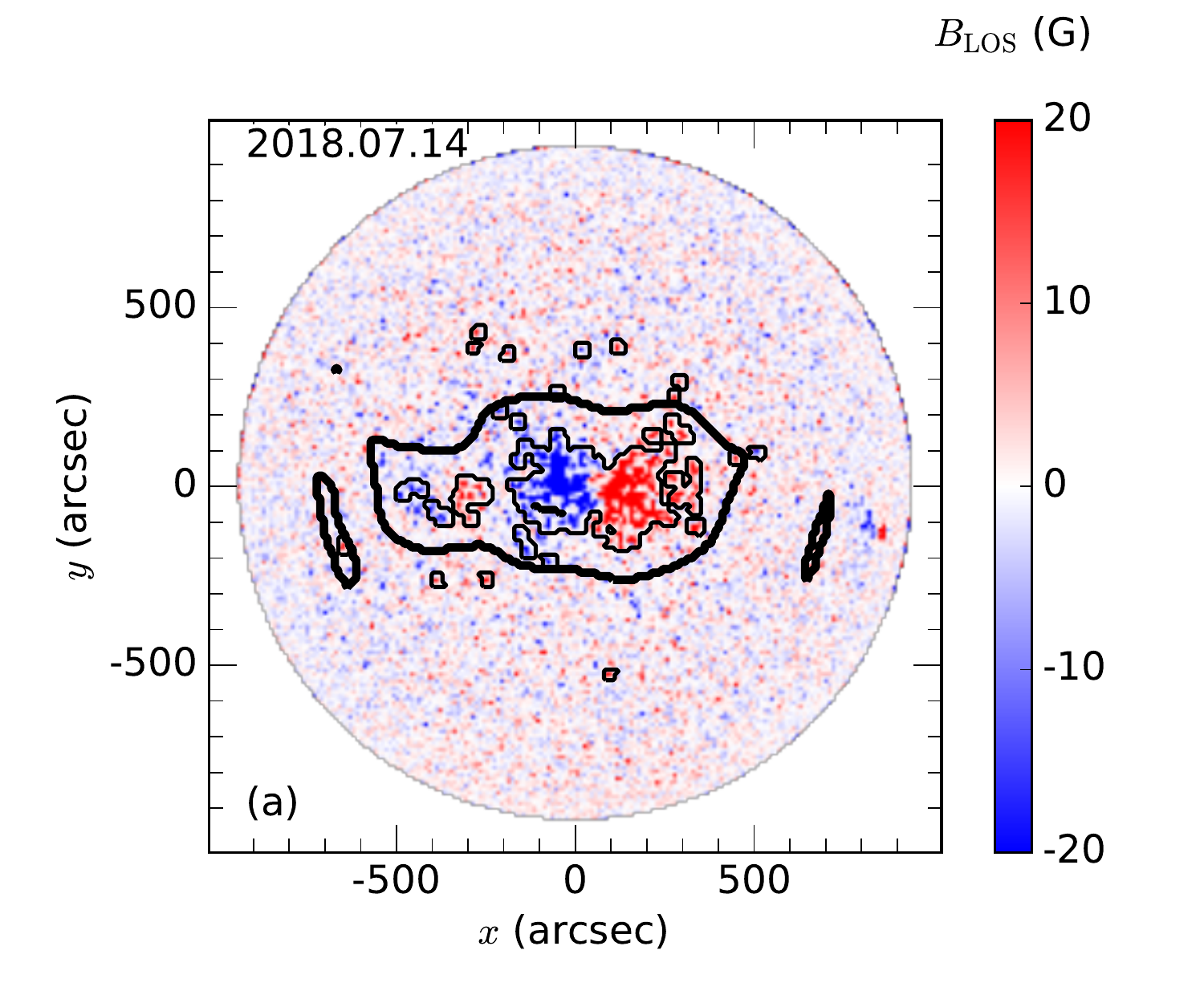}
    \includegraphics[width=0.5\textwidth]{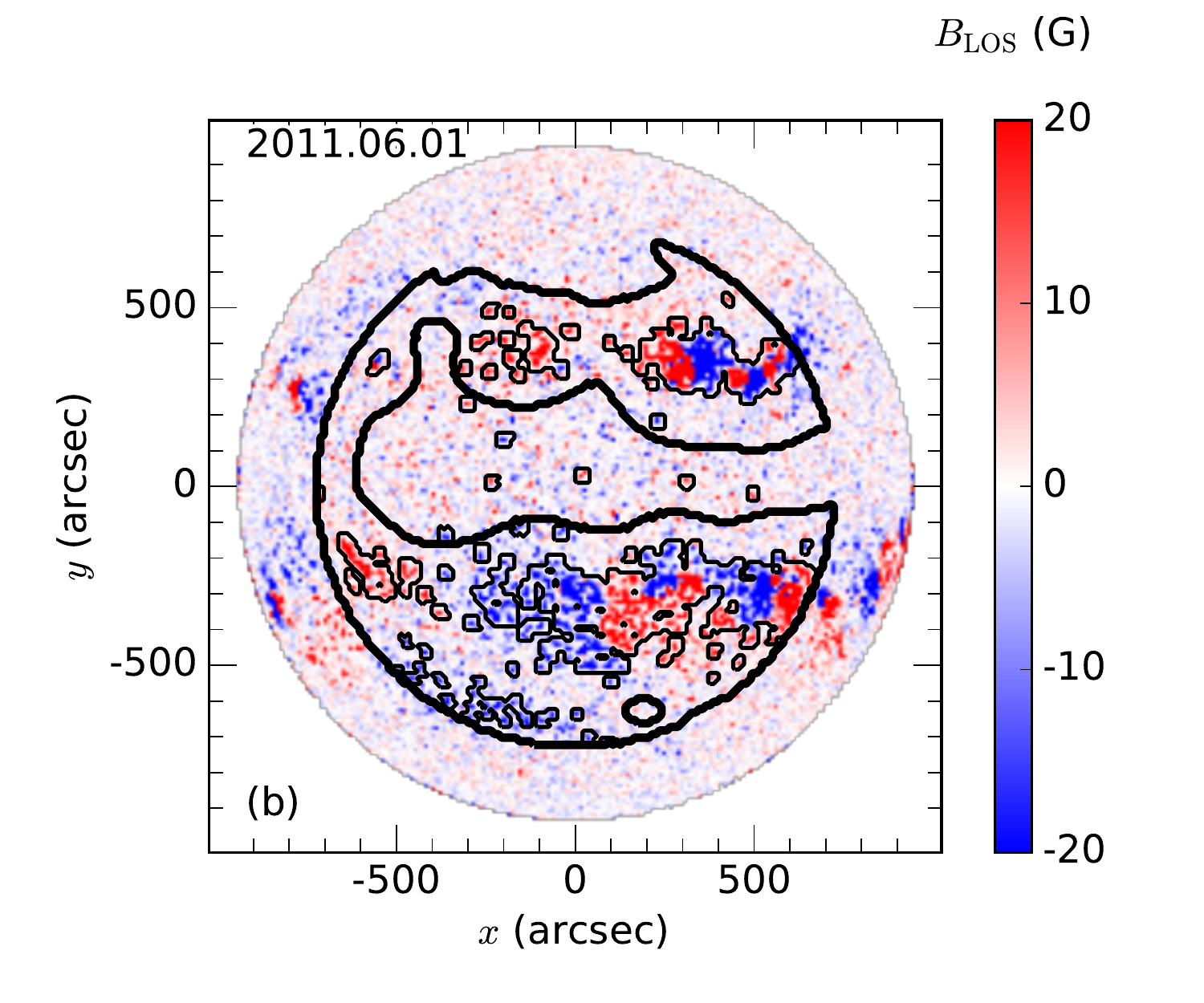}
    \caption{Contours around active-region flows (thick black lines) during solar minimum (\textit{panel a}) and during the rising phase (\textit{panel b}), superimposed on SDO/HMI magnetograms (line-of-sight magnetic field). The thin black lines represent the contours of the magnetic mask discussed in Section~\ref{sec:magmask}. We only consider pixels whose great-circle distance from disk center is less than $50\degr$, and hence the contours do not encompass the magnetic features near the limb.
    }
    \label{fig:fig1}
\end{figure}

We aim to determine smooth contours around the active-region flows based on a spatially-smoothed magnetic field.
We use a 2D Gaussian kernel to smooth $|B_r|$, with a full width at half maximum (FWHM) of $10\degr$ (here, $1\degr$ is $1$ heliographic degree) because it is the typical horizontal extent of the flows into active regions (see Fig.~\ref{fig:figa1}). We then define the contours around the active regions as the lines along which the smoothed $|B_r|$ is equal to a magnetic threshold $b$. We choose $b$ to be $3.5$~G (see Appendix \ref{sec:contour}). By aligning the centers of active regions with a total magnetic flux above $10^{21}$~Mx in a way similar to \citet{2019Braun}, we indeed find that this choice allows us to capture flows as far as $12^\circ$ away from the active region centers. The contours capture not only the active-region flows, but also the flows around the diffuse flux that is far less concentrated, as seen in Fig.~\ref{fig:fig1}.

\subsubsection{Background flows \label{sec:bkg_smooth}}

We consider the flows outside the contours as the background flows which consist of the potential systematics. We generate the background flow maps by computing monthly averages of the flows using only the pixels outside the contours.

The monthly background flow maps can contain little to no data at mid-latitudes during the solar maximum.
Therefore, 
we smooth in time with a Gaussian of FWHM of one year. This width is large enough that it smooths out the regions containing no data for several months, while the background is not expected to vary significantly on that time scale (see Appendix~\ref{sec:bkg_flows}). We further smooth in space with a 2D Gaussian of FWHM of ten pixels, corresponding to an angular distance of $6\degr$ at disk center.

These smoothed background flows exhibit a modulation on the time scale of a solar cycle. This modulation is consistent with what is described by \citet{2021Gottschling}, although we compute the background flows in a different way. They are studied in more detail in Appendix \ref{sec:bkg_flows}.

\subsection{Active-region flows \label{sec:ar_flows}}

The flows inside the contours are the superposition of the background flows and the active-region flows.
We therefore subtract the smoothed background flows from the 30-minute cadence flow maps; the remaining flows are the active-region flows.  The residual flows outside of the contours are not discarded after the subtraction of the background flows. They are expected to be random noise from convection.

We track these active-region flow maps at the Carrington rate on a daily basis using noon as a reference, and remap them using the Plate Carree projection into the heliographic coordinate system. The spatial resolution is $0.6^\circ$ per pixel. Following the formulas in the appendix of \citet{2017Loeptien}, we convert the velocities from pixels per second to meters per second. Then we average the flows daily. The colatitudinal and longitudinal components of these daily-averaged flows are denoted respectively by $v_{\theta}^{\mathrm{AR}}(\theta,\phi,t)$ and $v_{\phi}^{\mathrm{AR}}(\theta,\phi,t)$.

\section{Temporal variation of active-region flows \label{sec:flows}}
\begin{figure}[t]
    \centering
    \includegraphics[width=0.5\textwidth]{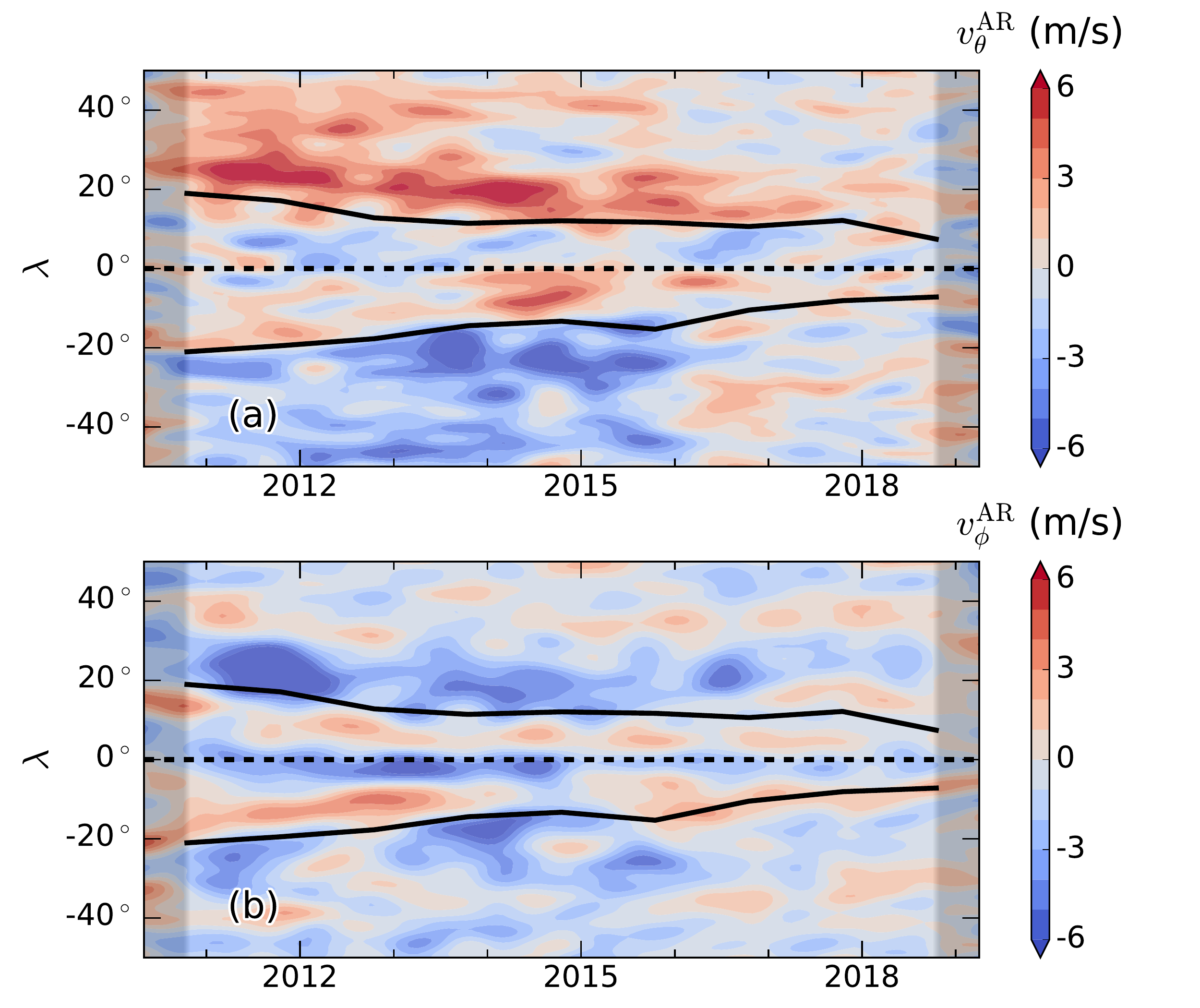}
    \caption{Longitudinal average of $v_{\theta}^{\mathrm{AR}}$ (\textit{panel a}; positive values are southward) and $v_{\phi}^{\mathrm{AR}}$ (\textit{panel b}; positive values are prograde), as a function of time and latitude. They were further smoothed in latitude with a Gaussian of FWHM of $3.6^\circ$ and in time with a Gaussian of FWHM of 6 months. The black lines show the mean active latitudes. The ticks on the horizontal axis indicate the beginning of each year. The shaded areas indicate the times when the edge effects become visible due to the one-year temporal smoothing of the background flows.}
    \label{fig:fig2}
\end{figure}
\begin{figure}
    \centering
    \includegraphics[width=0.45\textwidth,height=0.90\textheight]{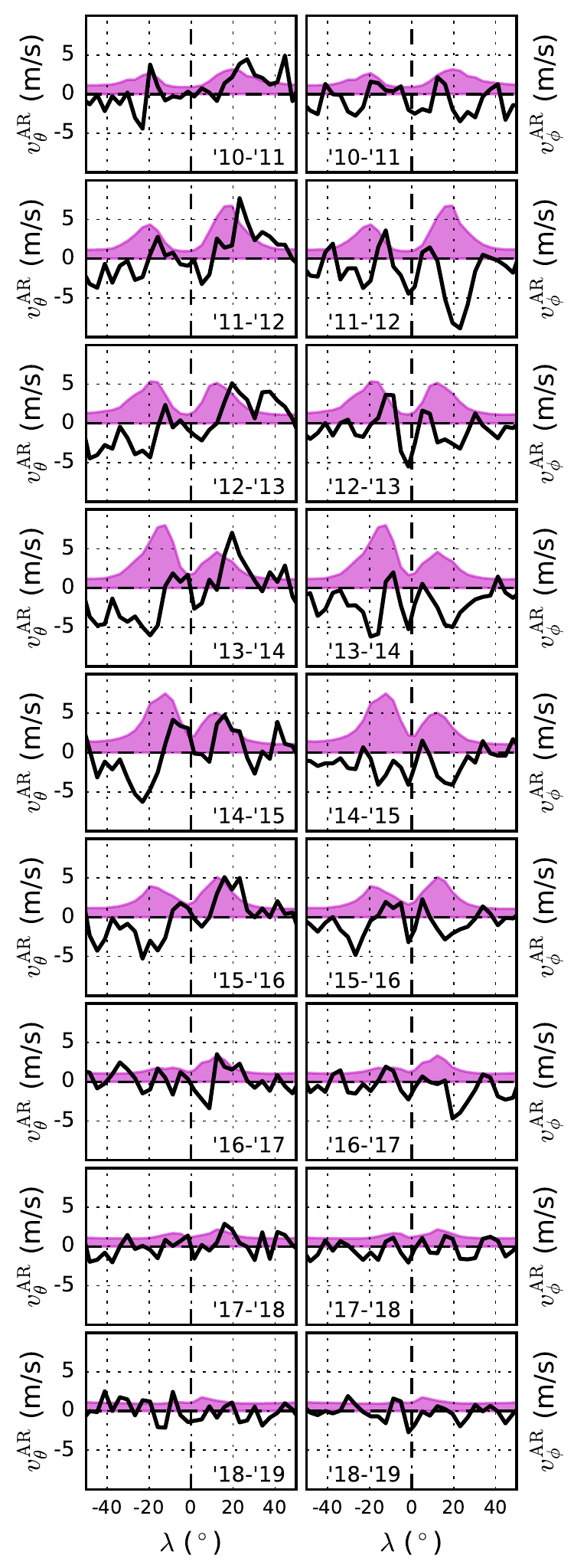}
    \caption{Longitudinally- and yearly-averaged $v_{\theta}^{\mathrm{AR}}$ and $v_{\phi}^{\mathrm{AR}}$ (black lines),
    binned every $3.6\degr$ in latitude.
    The standard error of the mean in each binning interval is about $1$~m/s for $v_\theta^\mathrm{AR}$ and $0.4$~m/s for $v_\phi^\mathrm{AR}$, and is not shown here. The magenta filling indicates the distribution of $|B_r|$ (normalized to the same arbitrary constant in all time periods), which was averaged and binned in the same way as the flows.}
    \label{fig:fig3}
\end{figure}

To examine the temporal and latitudinal evolution of the active-region flows, we average $v_{\theta}^{\mathrm{AR}}$ and $v_{\phi}^{\mathrm{AR}}$ in longitude within $\pm 15^\circ$ of the central meridian.
Figure~\ref{fig:fig2} shows the longitudinal averages of $v_{\theta}^{\mathrm{AR}}$ and $v_{\phi}^{\mathrm{AR}}$ as a function of time and latitude (denoted as $\lambda=90\degr-\theta$). Figure~\ref{fig:fig3} shows the yearly-averaged flows as a function of latitude, together with the magnetic activity.

Figure~\ref{fig:fig2}a shows an inflow pattern converging toward the mean latitude of activity in both hemispheres. The inflows
are stronger and cover a greater latitudinal range on the poleward side of the active latitudes. We note that, in \citet{2019Braun} and \citet{2021Gottschling}, the inflows are mostly symmetric with respect to the center of the active regions. One possible reason is that they use an ensemble averaging over active regions whose centers have been aligned with each other, while we use a longitudinal long-term averaging that
makes the inflows on the equatorward side of the northern and southern active latitudes partly cancel each other out. 
In the left column of Fig.~\ref{fig:fig3}, the amplitude of the yearly-averaged $v_{\theta}^{\mathrm{AR}}$ reaches extrema of over $7$~m/s in the north in 2011\,--\,2012 and over $6$~m/s in the south in 2014\,--\,2015.
Likewise, the solar magnetic activity peaks in 2011 in the northern hemisphere and in 2014 in the southern hemisphere.
Therefore the amplitude of the inflows is clearly correlated with the strength of the solar activity. The amplitude is consistent with that from, e.g., \citet{2003Gizon}, \citet{2008GonzalezHernandez}, \citet{2020Komm}, who isolated regions of magnetic activity and their surroundings in a way similar to what we do here.
We note that \citet{2003Spruit} interpreted the inflows as a consequence of the enhanced cooling in the magnetic regions and predicted inflows of $\sim$6~m/s toward the activity belts. In the bottom two panels, $v_\theta^\mathrm{AR}$ shows no active-region flows but rather random oscillations that have the same magnitude at all latitudes, as expected toward the end of the solar cycle.

Figure~\ref{fig:fig2}b shows that the toroidal component consists of a retrograde flow (with respect to the background flow) on the poleward side of the active latitudes, and a generally prograde flow on the equatorward side. This pattern may be consistent with the results of \citet{2020Komm}.
It is known that the torsional oscillations, which are the time-varying part of the solar rotation, exhibit a shear flow around the active latitudes, with a faster-rotating band on the equatorward side and a slower-rotating one on the poleward side \citep[e.g.,][]{1980Howard}. Therefore, the toroidal component of the active-region flows seen here contributes to the torsional oscillations to some extent.

The right column of Fig.~\ref{fig:fig3} shows that the amplitude of the toroidal component is also correlated with the amplitude of the solar magnetic activity. The retrograde flow on the poleward side of the active latitudes reaches over $8$~m/s in the north in 2011\,--\,2012 and over $6$~m/s in the south in 2013\,--\,2014. We note that \citet{2019Braun} and \citet{2021Gottschling} reported the presence of a retrograde flow surrounding active regions preferentially on the poleward side. In addition, there is a weaker prograde flow on the equatorward side of the activity belts. The presence of this signal might be consistent with the idea of a cyclonic circulation around active regions that derives from the model of \citet{2003Spruit} and that was also described by \cite{2009Hindman}.

\section{Forward-modeled helioseismic travel-time perturbations \label{sec:traveltimes}}

\subsection{Computation of travel-time perturbations}

\begin{figure}[t]
    \centering
    \includegraphics[width=0.5\textwidth]{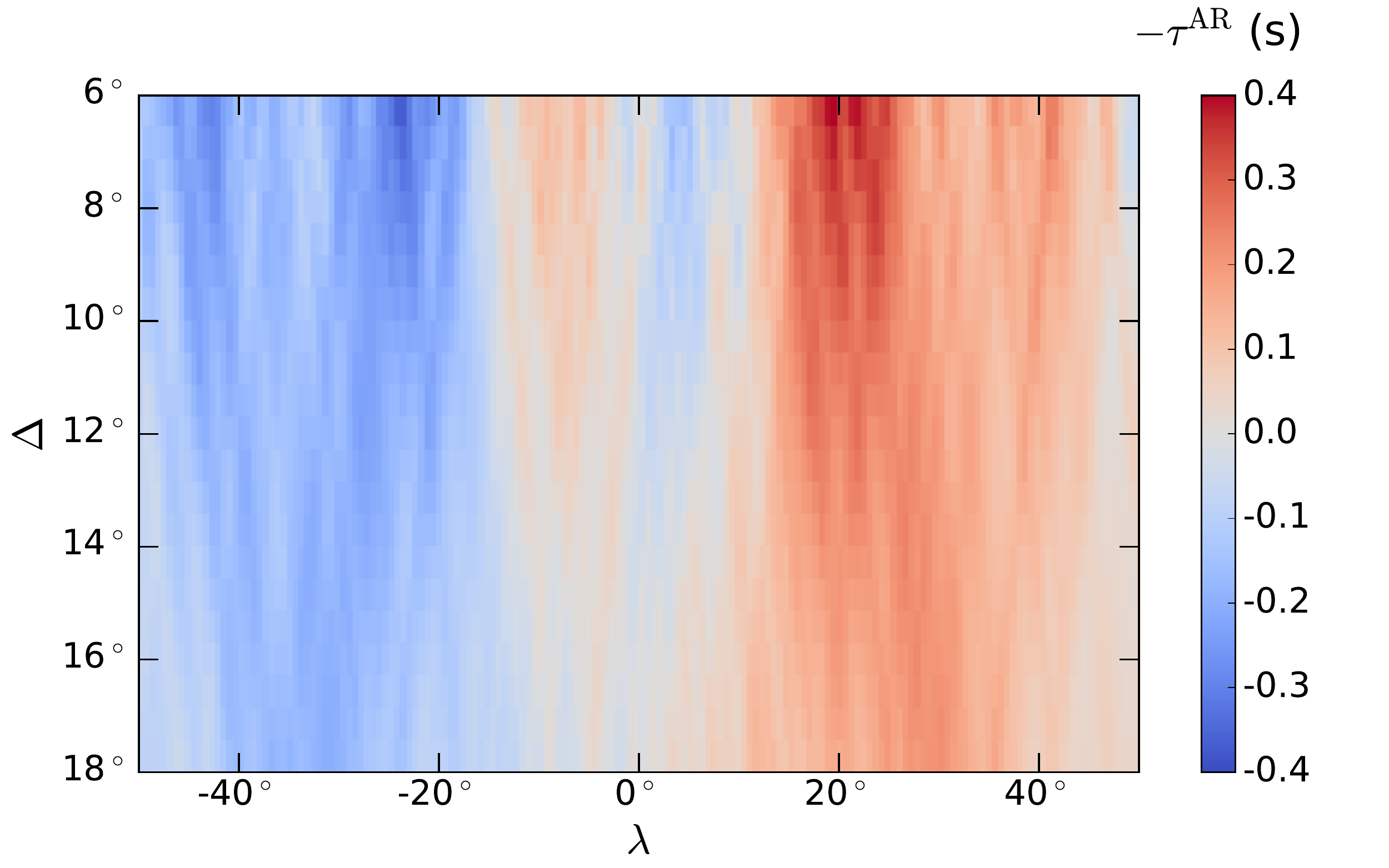}
    \caption{Forward travel-time perturbations $-\tau^{\mathrm{AR}}$, as a function of latitude and separation distance. They have been averaged in longitude around the central meridian and in time from January 2011 to December 2014.}
    \label{fig:fig4}
\end{figure}

\begin{figure*}[t]
    \centering
    \includegraphics[width=\textwidth]{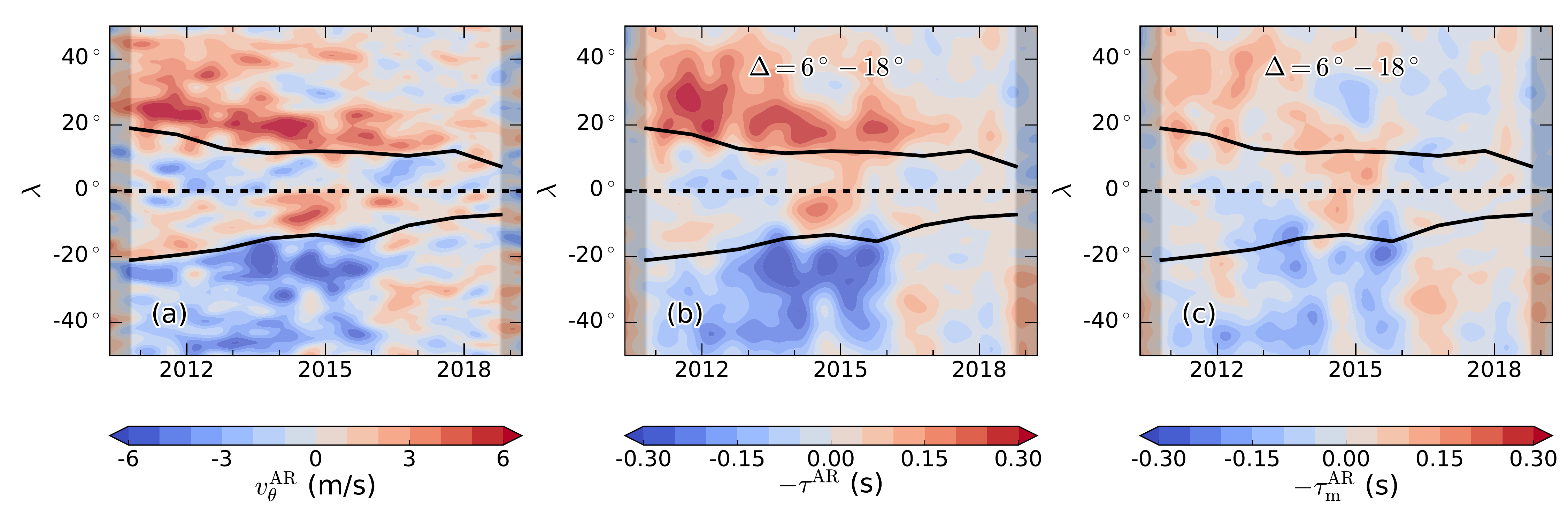}
    \includegraphics[width=\textwidth]{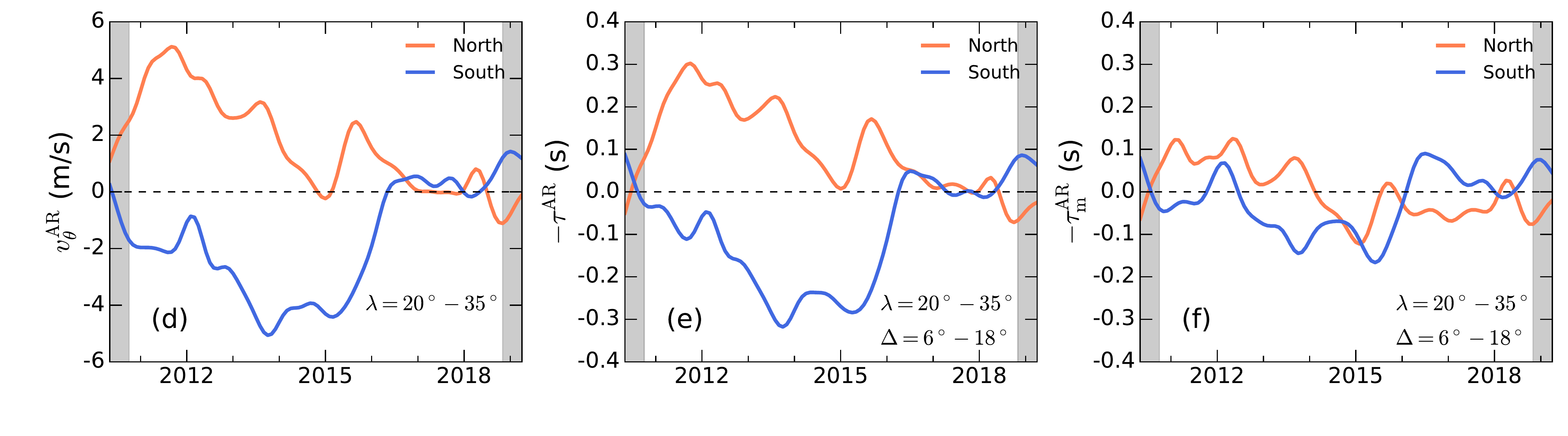}
    \caption{\textit{Panel a}: Longitudinally-averaged active-region flows $v_{\theta}^{\mathrm{AR}}$ as plotted in Fig.~\ref{fig:fig2}, shown again here for convenience of comparison. \textit{Panel b}: Longitudinally-averaged forward travel-time perturbations $-\tau^{\mathrm{AR}}$, further averaged over the separation distances $6\degr$\,--\,$18\degr$. \textit{Panel c}: same as \textit{panel b}, but for the travel-time perturbations with the magnetic mask applied, $-\tau_\mathrm{m}^\mathrm{AR}$ (discussed in Section~\ref{sec:magmask}). We applied the same smoothing as in Fig.~\ref{fig:fig2}. The black lines show the mean active latitudes. \textit{Panels d--f}: Averages of the top panels over the latitudes $\lambda=20^\circ-35^\circ$ in each hemisphere. The typical errors in the flows over the latitudinal range of interest are about $0.6$~m/s.
    For the travel-time perturbations, the errors are in the range of $0.002$\,--\,$0.008$~s, with the largest error for $\tau_\mathrm{m}^\mathrm{AR}$ during the peaks of activity. The errors are not shown here. In all panels, the shaded areas indicate the times when the edge effects become visible due to the one-year temporal smoothing of the background flows.}
    \label{fig:fig5}
\end{figure*}

We compute the forward travel-time perturbations associated with the active-region flows to compare with the solar-cycle variations of the helioseismic measurements of the meridional circulation. To this end, we first construct a 3D flow profile associated with the active-region flows at the surface. We neglect the contribution from the vertical flow to the travel times. For the colatitudinal and longitudinal components $u_\theta$ and $u_\phi$, we choose a constant profile with depth such that
\begin{align}
    & u_\theta(\vect{r},t) = v_\theta^{\mathrm{AR}}(\theta,\phi,t), \\
    & u_\phi(\vect{r},t) = v_\phi^{\mathrm{AR}}(\theta,\phi,t),
\end{align}
where $\vect{r}=(r,\theta,\phi)$ with $r$ being the distance to the solar center. Although it would be more realistic to choose a flow profile where the amplitude decreases with depth, choosing a constant profile allows us to estimate an upper bound for the amplitude of the travel-time perturbations due to active-region flows. 

In order to directly compare with \citet{2020Gizon}, we use an arc-to-arc geometry, with the travel-time perturbations being computed between pairs of points placed on two opposite arcs. The paired arcs, each subtending an angle of $30^\circ$, are aligned in the north-south direction as described by \citet{2017Liang}. 
Let's denote $\psi$ the angle between a meridian and the ray path connecting the paired points on the arcs, $(\theta_0,\phi_0)$ the mid-point between the paired arcs, and $\Delta$ the separation distance between the paired points. The travel-time perturbation is thus defined as \citep[e.g.,][]{2017Gizon,2018Fournier}
\begin{equation}
\begin{split}
    \tau
    (\theta_0,\phi_0,\Delta,\psi,t)=\int_\odot &
   \Bigl( {K}_\theta ({\vect{r}} ; \theta_0, \phi_0, \Delta,\psi) \; u_\theta(\vect{r}, t) \\
  & +{K}_\phi ({\vect{r}} ; \theta_0, \phi_0, \Delta,\psi) \; u_\phi(\vect{r}, t)\Bigl)\;  \dd\vect{r},
\end{split}
\end{equation}
where the integral is taken over the whole volume of the Sun, and $K_\theta$ and $K_\phi$ are the colatitudinal and longitudinal components of the sensitivity kernel (see Appendix \ref{sec:born}). Here, by convention, a northward flow perturbation corresponds to a positive travel-time perturbation for short separation distances.

The forward travel-time perturbation is averaged in the same way as in \citet{2020Gizon}; that is,
\begin{equation}
    \tau^{\mathrm{AR}}(\theta_0,\Delta,t) =  \left(N_{\phi_0}N_\psi \right)^{-1} \times
    \sum\limits_{\phi_0,\psi} \; 
    \tau
    (\theta_0,\phi_0,\Delta,\psi,t),
    \label{eq:av_tau}
\end{equation}
where the sum over $\phi_0$ is taken within $\pm15\degr$, the sum over $\psi$ is taken within $\pm15\degr$, $N_{\phi_0}$ is the number of points used in the longitudinal average, and $N_\psi$ is the number of points on an arc for each distance.

Figure~\ref{fig:fig4} shows the forward travel-time perturbations as a function of latitude and separation distance after we averaged over the active period from January 2011 to December 2014. Even though we chose a constant flow profile with depth, the forward travel-time perturbations could still change sign with increasing separation distance, for example, when the separation distance is larger than the spatial scale of active region flows. Our $\tau^\mathrm{AR}$ decreases with separation distance but does not change sign up to at least $\Delta=18^\circ$, which is similar to the modeled travel-time perturbations from \citet{2018Liang}.

We show the longitudinally-averaged $v_{\theta}^{\mathrm{AR}}$ (as plotted in Fig. \ref{fig:fig2}) in Fig.~\ref{fig:fig5}a, and the corresponding $\tau^{\mathrm{AR}}$ averaged over the separation distances $6^\circ$\,--\,$18^\circ$ in Fig.~\ref{fig:fig5}b. We can clearly see the inflow pattern converging toward the active latitudes in Fig.~\ref{fig:fig5}b, with an amplitude that varies throughout the solar cycle as in panel Fig.~\ref{fig:fig5}a. When averaged over the latitudes $20^\circ-35^\circ$ in each hemisphere, the amplitude of the inflows reaches $5$~m/s in 2011 in the north and $-5$~m/s in 2014 in the south (Fig.~\ref{fig:fig5}d). Similarly, the amplitude of $\tau^\mathrm{AR}$ reaches extrema of $0.3$~s during these active periods (Fig.~\ref{fig:fig5}e). 
For comparison, we also computed $\tau^{\mathrm{AR}}$ using the radial flow profile from the shallow model LC2 described in \citet{2018Liang};
in this case, the amplitude of $\tau^{\mathrm{AR}}$ reaches extrema of $0.15$~s. 

\subsection{Effect of masking travel-time perturbations inside magnetic regions \label{sec:magmask}}

\citet{2020Gizon} excluded the travel-time perturbations measured within magnetic regions from the averages, as those measurements introduce a systematic error that resembles a divergent flow pattern \citep{2015Liang}. In order to compare with their results, we apply the same masking as they did. The contours of the mask are shown as the thin black lines in Fig.~\ref{fig:fig1}. We rewrite the forward travel-time perturbation averaging with a weighting function $w$ that is equal to zero if the forward travel-time perturbation is excluded and one elsewhere. Equation~\ref{eq:av_tau} then becomes
\begin{equation}
\begin{split}
    \tau^{\mathrm{AR}}_\mathrm{m}(\theta_0,\Delta,t) &=  \left(\sum\limits_{\phi_0,\psi}w(\theta_0,\phi_0,\Delta,\psi,t) \right)^{-1} \times \\
    & \sum\limits_{\phi_0,\psi} \; w(\theta_0,\phi_0,\Delta,\psi,t)\;
    \tau
    (\theta_0,\phi_0,\Delta,\psi,t).
\end{split}
\label{eq:av_tau_bis}
\end{equation}

\begin{figure}[t]
    \centering
    \includegraphics[width=0.5\textwidth]{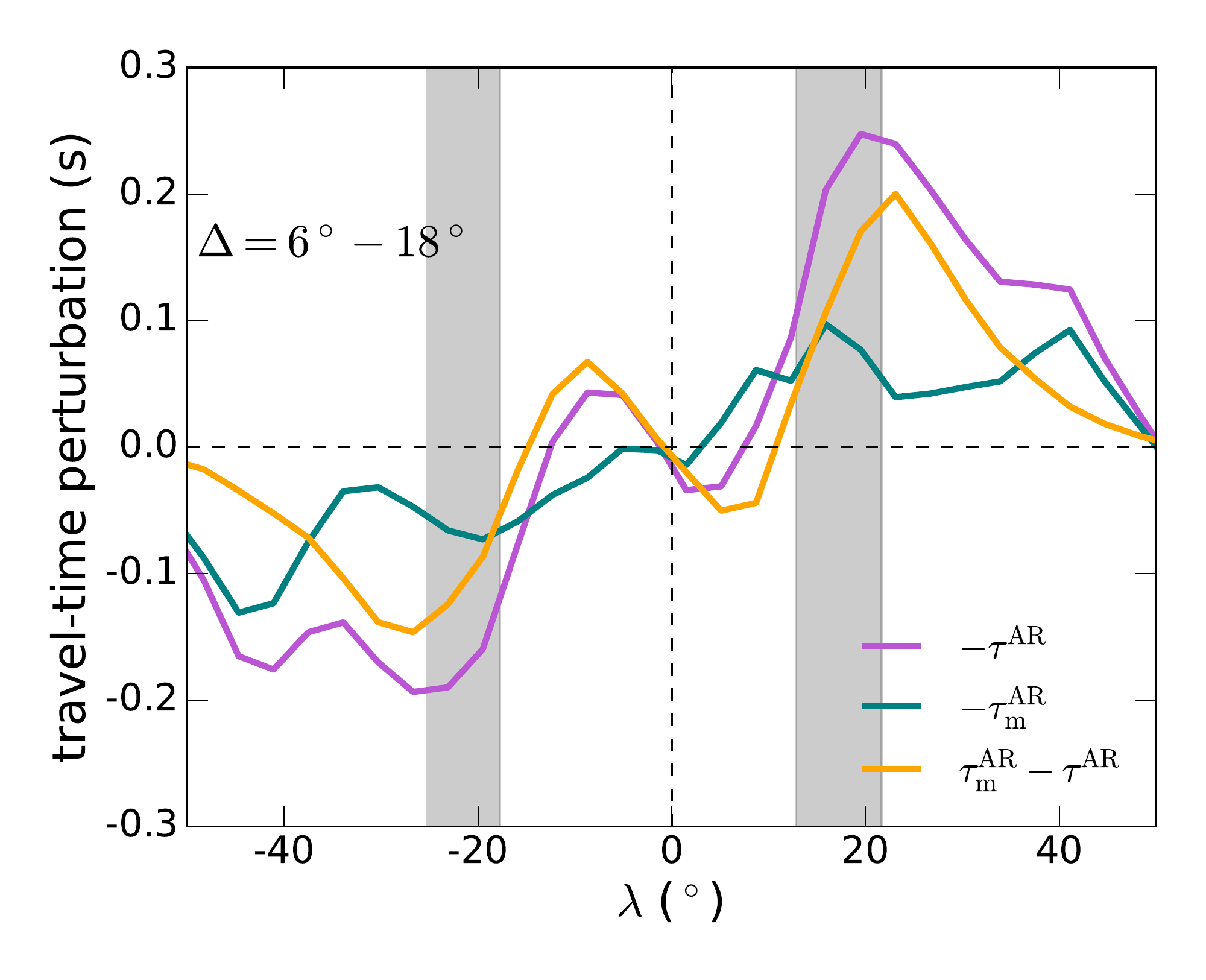}
    \caption{Longitudinally-averaged forward travel-time perturbations, further averaged over the separation distances $6\degr$\,--\,$18\degr$ and over the time period from January 2011 to December 2014, as a function of latitude. They have been binned every $3.6\degr$ in latitude. The typical standard error of the mean, computed in each binning interval, is about $0.003$~s, and is not shown here. The vertical gray shaded areas show the mean active latitudes.}
    \label{fig:fig6}
\end{figure}

The third column of Fig.~\ref{fig:fig5} shows the forward travel-time perturbations after the magnetic mask is applied, denoted by $\tau_\mathrm{m}^\mathrm{AR}$. These perturbations are associated with the flows located 
outside the thin lines and inside the thick lines in Fig.~\ref{fig:fig1}. 
The $\tau_\mathrm{m}^\mathrm{AR}$ in Fig.~\ref{fig:fig5}c resembles the $\tau^\mathrm{AR}$ in Fig.~\ref{fig:fig5}b but the amplitude of the inflow pattern is strongly reduced during solar maximum.
Figure~\ref{fig:fig5}f shows that, when averaged over mid-latitudes, the amplitude reaches extrema in 2011\,--\,2012 in the north and in 2014\,--\,2015 in the south; the extrema are three times smaller than that of $\tau^\mathrm{AR}$ in Fig.~\ref{fig:fig5}e.

Figure~\ref{fig:fig6} presents a comparison between $\tau^\mathrm{AR}$ and $\tau_\mathrm{m}^\mathrm{AR}$, averaged over four active years from January 2011 to December 2014.
$\tau^{\mathrm{AR}}$ shows an inflow pattern with an amplitude reaching extrema at a latitude of about $20\degr$ in both hemispheres. With the masking, the inflow pattern in $\tau_\mathrm{m}^\mathrm{AR}$ converging toward the active latitudes is nearly gone.
The difference $\tau_\mathrm{m}^\mathrm{AR}-\tau^\mathrm{AR}$ presents a clear inflow pattern toward the active latitudes, comparable to $\tau^\mathrm{AR}$. Therefore, the masking of the magnetic regions removes most of the inflow pattern from the averaged travel-time perturbations.

We note that the masking of magnetic pixels implemented in \cite{2020Gizon} checks an area of $4\times4$ pixels around either of the paired foot points in the arc-to-arc geometry \citep{2017Liang}; if the field strength is greater than a threshold in this area, the paired points are excluded from the averages.
We found that if only the nearest pixel to the foot point (instead of a $4\times4$-pixel area) is checked, the reduction of the inflow pattern in the forward travel-time perturbations is not as strong as the aforementioned results.

\section{Extension to May 1996\,--\,April 2019 \label{sec:extension}}

\begin{figure}[t]
    \centering
    \includegraphics[width=0.5\textwidth]{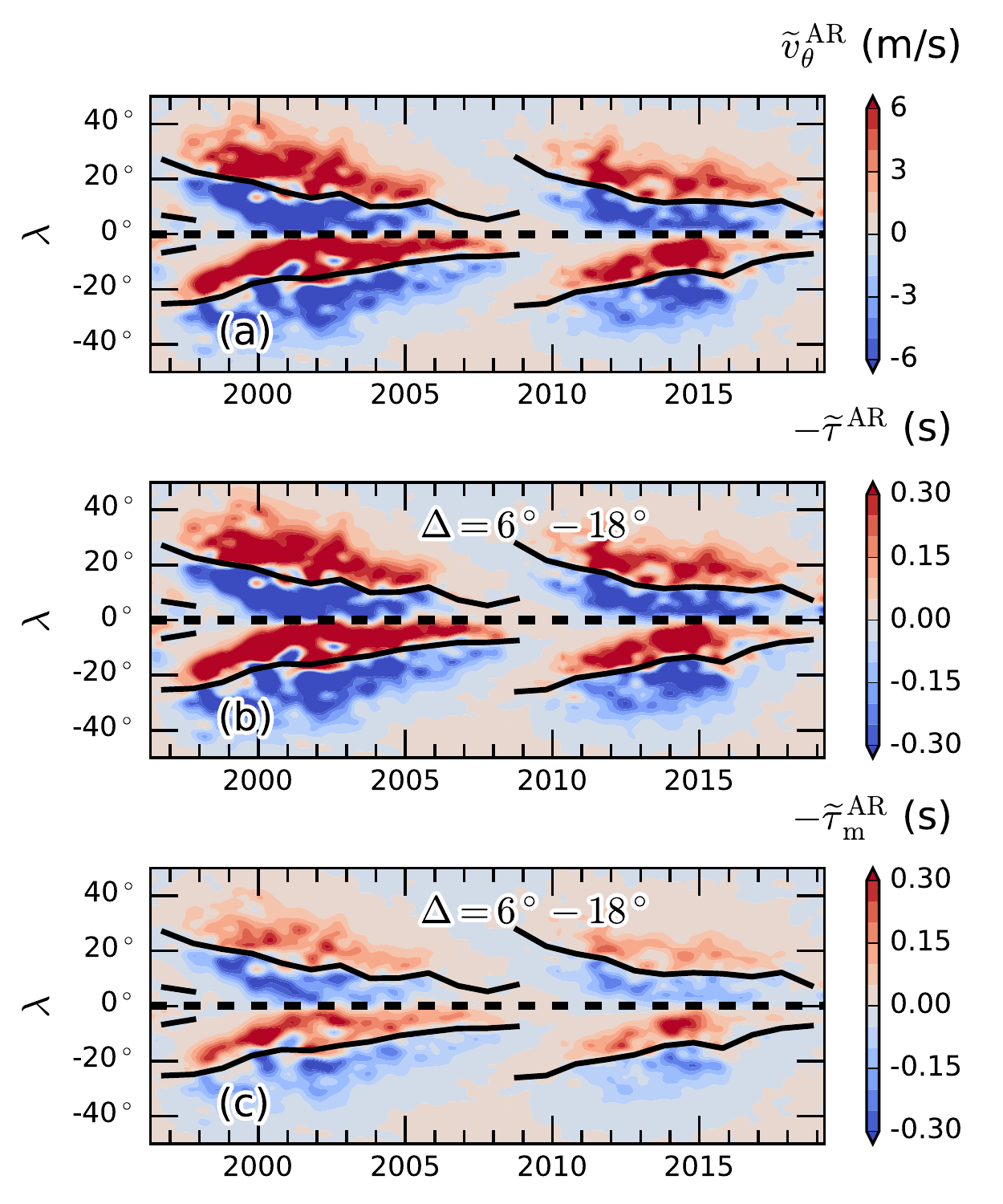}
    \caption{Modeled inflows $\widetilde v_\theta^\mathrm{\,AR}$ (\textit{panel a}), and travel-time perturbations without magnetic mask $-\widetilde \tau^\mathrm{\,AR}$ (\textit{panel b}) and with magnetic mask $-\widetilde \tau_\mathrm{m}^\mathrm{\,AR}$ (\textit{panel c}) from May 1996 to April 2019.
    The conversion constant for producing $\widetilde \tau^\mathrm{\,AR}$ and $\widetilde \tau_\mathrm{m}^\mathrm{\,AR}$ is derived from the HMI data averaged over the separation distances $6\degr$\,--\,$18\degr$.
    The same smoothing as in Fig.~\ref{fig:fig2} is applied. The black lines show the mean active latitudes.}
    \label{fig:fig7}
\end{figure}
\begin{figure}[h!]
    \centering
    \includegraphics[width=0.5\textwidth]{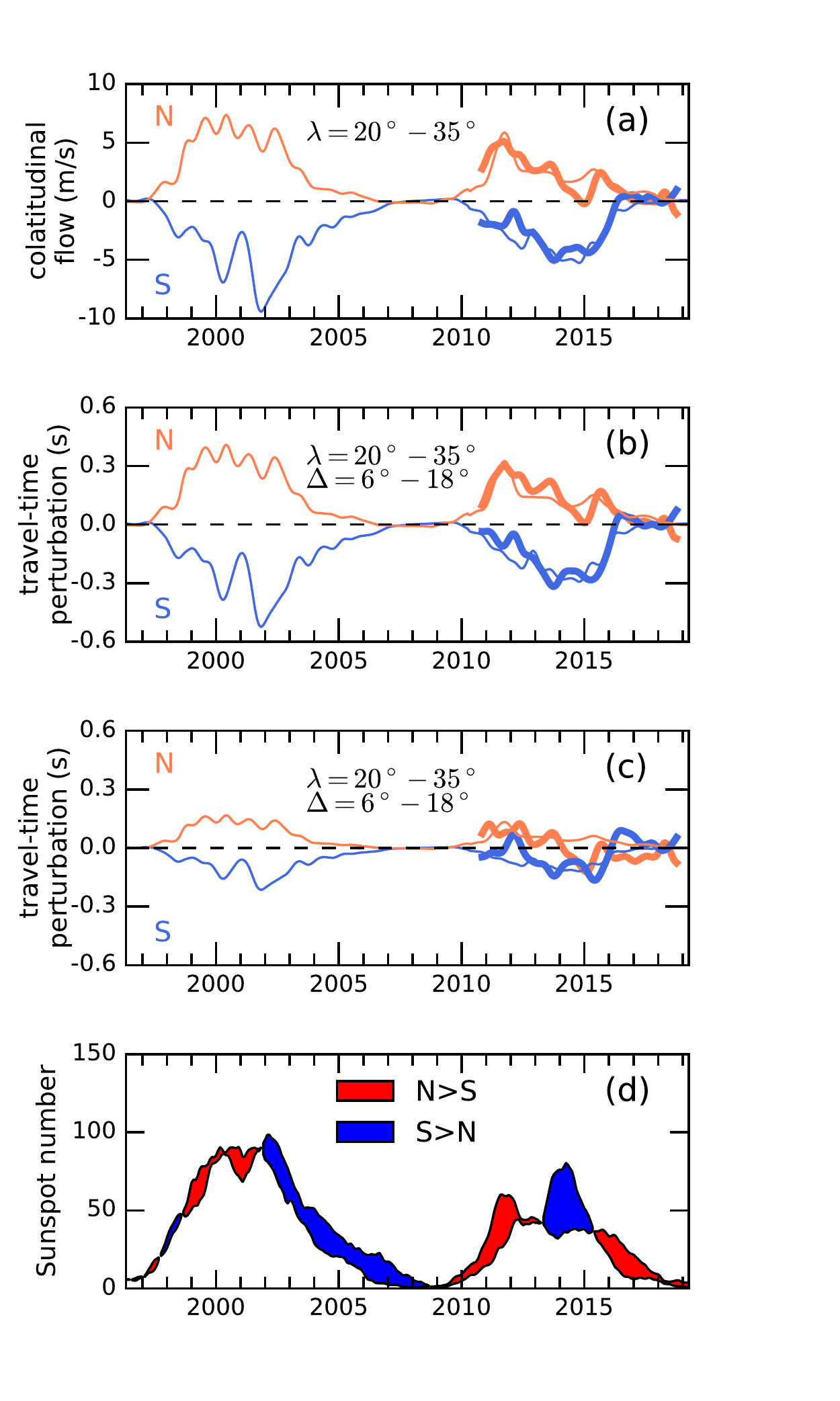}
    \caption{\textit{Panels a--c}: Averages of $\widetilde v_\theta^\mathrm{\,AR}$, $-\widetilde \tau^\mathrm{\,AR}$, and $-\widetilde \tau_\mathrm{m}^\mathrm{\,AR}$ from Figs.~\ref{fig:fig7}a--c over the latitudes $\lambda=20^\circ-35^\circ$ in each hemisphere (thin lines).
    The errors on the models, estimated from the misfits, are about $0.6$~m/s for \textit{panel a}, $0.04$~s for \textit{panel b}, and $0.03$~s for \textit{panel c}; they are not shown here.
    For comparison, the corresponding $v_\theta^\mathrm{AR}$, $-\tau^\mathrm{AR}$, and $-\tau_\mathrm{m}^\mathrm{AR}$ from Figs.~\ref{fig:fig5}d--f are overplotted in thick lines; the data in the first and last six months are not shown because the edge effects become visible due to the one-year temporal smoothing of the background flows. \textit{Panel d}: 13-month running mean of the hemispheric monthly sunspot numbers. The red (respectively blue) filling indicates an excess of sunspot numbers in the northern (respectively southern) hemisphere.
    }
    \label{fig:fig8}
\end{figure}
We aim to extend the analysis to cover the time period 1996\,--\,2019 in order to compare the results with \citet{2020Gizon}.
Although the LCT data are only available after May 2010 as they were computed from full-resolution HMI intensity images, we could use the magnetic field as a proxy for the active-region inflows from May 1996 to April 2010.
The correlation between the magnetic field and the inflows is visible in Fig.~\ref{fig:fig3}: the greater the latitudinal gradient of the magnetic field strength, the greater the inflows. Several models already exist in the dynamo literature to generate artificial active-region inflows based on the observations of the magnetic field \citep[e.g.,][]{2006DeRosa,2010Cameron,2012Cameron}.

We minimize the cost function
\begin{equation}
    \sum_{\theta,t} \left|\langle v_\theta^{AR}(\theta,\phi,t) \rangle-c_0 \frac{\partial \langle |B_r(\theta,\phi,t)| \rangle}{\partial \theta} \right|^2
\end{equation}
to determine the constant of proportionality $c_0$. Here
$\langle \, \cdot \, \rangle$ denotes the averaging in longitude within $\pm 15\degr$ of the central meridian and the smoothing in latitude and time in the way described in Fig.~\ref{fig:fig2}. We use only the data from January 2011 to December 2014 and in the latitudinal range $20\degr$\,--\,$35\degr$, where the inflows are the strongest. We find $c_0\simeq 0.085$~m~s$^{-1}$~G$^{-1}$.

We compute the modeled inflows $\widetilde v_\theta^\mathrm{\,AR}$ using the SOHO/MDI magnetograms \citep{1995Scherrer} from May 1996 to April 2010 and the HMI magnetograms from May 2010 to April 2019.
The line-of-sight magnetic field inferred from HMI data is smaller than that from MDI data \citep{2012Liu}; the scaling factor depends on the location on the disk and on the field strength. We determine this scaling factor using $B_r$ at latitudes $20\degr$\,--\,$35\degr$ in each hemisphere and longitudes within $\pm 15\degr$ of the central meridian, during the time period from May 2010 to April 2011 when both data sets are available.
We find the scaling factor is about $0.74$ which is then applied to the MDI data.

Finally, based on the similarity between the inflows and the travel-time perturbations shown in Fig. \ref{fig:fig5},
we use a conversion constant to convert from the modeled flows $\widetilde v_\theta^\mathrm{\,AR}$ to modeled travel-time perturbations from May 1996 to April 2019. For an average over the separation distances $6^\circ$\,--\,$18^\circ$, we find the conversion constant to be $-0.056$~s$^2$/m for the case without the magnetic mask $\widetilde \tau^\mathrm{\,AR}$ and $-0.023$~s$^2$/m for the case with the magnetic mask $\widetilde{\tau}_\mathrm{m}^\mathrm{\,AR}$, where the tilde indicates the modeled quantities. We present the scatter plots of the data used for the modeling in Appendix~\ref{sec:modeling}.

Figure~\ref{fig:fig7} shows the modeled inflows and the travel-time perturbations without and with magnetic mask as a function of time and latitude, over Cycle 23 (1996\,--\,2008) and Cycle 24 (2008\,--\,2019). All the panels exhibit the inflow pattern with the flows generally converging toward the mean active latitudes, which is similar to that in Fig.~\ref{fig:fig5}. The amplitude of the models is comparable to that of the observations for both the inflows and the travel-time perturbations on the poleward side of the activity belts. However, on the equatorward side, the amplitude of the models is greater because we determined the proportionality constant $c_0$ using only the latitudes $20\degr$\,--\,$35\degr$ in each hemisphere.

Figure~\ref{fig:fig8} shows how the models compare with the observations when averaged over the latitudes $20^\circ$\,--\,$35^\circ$ in each hemisphere. $\widetilde v_\theta^\mathrm{\,AR}$ (Fig.~\ref{fig:fig8}a) and $\widetilde \tau^\mathrm{\,AR}$ (Fig.~\ref{fig:fig8}b) generally match $v_\theta^\mathrm{AR}$ and $\tau^\mathrm{AR}$ well in both hemispheres, especially during the peaks of magnetic activity. When the magnetic mask is applied (Fig.~\ref{fig:fig8}c), the model $\widetilde \tau_\mathrm{m}^\mathrm{\,AR}$ retrieves the correct order of magnitude but does not fully capture $\tau_\mathrm{m}^\mathrm{AR}$. Furthermore, the amplitude of the models is greater during the Cycle 23 solar maximum than that during the Cycle 24 solar maximum, which is consistent with the sunspot number presented in Fig.~\ref{fig:fig8}d.

\section{Comparison with helioseismic measurements \label{sec:comparison}}

\begin{figure}[t]
    \centering
    \includegraphics[width=0.5\textwidth]{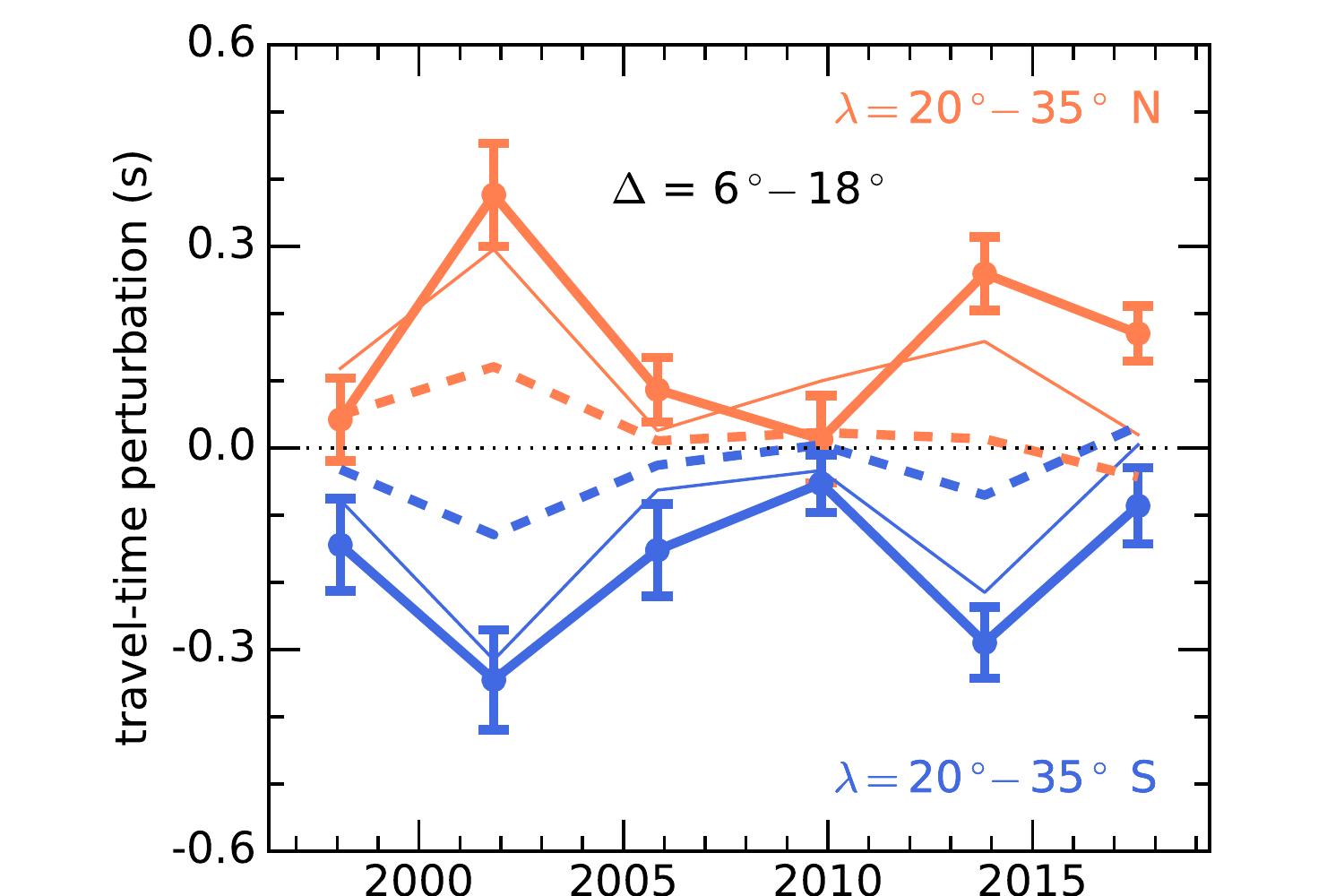}
    \caption{
    Comparison between the north-south travel-time perturbations as measured by \citet{2020Gizon} (thick solid lines) and the forward travel-time perturbations without (thin solid line) and with (thick dashed line) the magnetic mask. The measurements were averaged over the latitudes $\lambda=20\degr$\,--\,$35\degr$ in each hemisphere, over the separation distances $\Delta=6\degr$\,--\,$18\degr$, and over consecutive 4-year time intervals. The average over the quiet-Sun period January 2007 to April 2009 was subtracted from the measurements, and the sign was reversed for the purpose of comparison. The corresponding $-\widetilde \tau^\mathrm{\,AR}$ and $-\widetilde \tau_\mathrm{m}^\mathrm{\,AR}$ are used from May 1996 to April 2010, and $- \tau^\mathrm{AR}$ and $- \tau_\mathrm{m}^\mathrm{AR}$ from May 2010 to April 2019.
    For the measurements, the error bars represent the standard error of the mean computed over the latitudinal range of interest. For the forward travel-time perturbations, the standard error of the mean is in the range of $0.001$\,--\,$0.003$~s, and is not shown here. For the modeled travel-time perturbations, the errors, not shown here, are about $0.013$~s for $-\widetilde \tau^\mathrm{\,AR}$ and $0.009$~s for $-\widetilde \tau_\mathrm{m}^\mathrm{\,AR}$.}
    \label{fig:fig9}
\end{figure}

We compare in Fig.~\ref{fig:fig9} our forward travel-time perturbations with the measurements from \citet{2020Gizon} spanning Cycles 23 and 24. \citet{2020Gizon} used the data from MDI and GONG \citep{1996Harvey} for the periods from May 1996 to April 2003 and from May 2003 to April 2019, respectively; the travel times associated with the magnetic pixels were excluded in their measurements. For our travel-time perturbations, 
we used $\widetilde \tau^\mathrm{\,AR}$ and $\widetilde \tau_\mathrm{m}^\mathrm{\,AR}$ for the period from May 1996 to April 2010, and $\tau^\mathrm{AR}$ and $\tau_\mathrm{m}^\mathrm{AR}$ for the period from May 2010 to April 2019.

The measurements of \citet{2020Gizon} exhibit a modulation in amplitude over the solar cycles that is correlated with the magnetic activity.
The $\tau^\mathrm{AR}$ and $\widetilde \tau^\mathrm{\,AR}$ also exhibit this modulation.
We remind the reader that we used a constant profile with depth for the active-region flow model to place an upper limit; that is, the magnitude of $\tau^\mathrm{AR}$ and $\widetilde \tau^\mathrm{\,AR}$ is overestimated. When the masking is taken into account, the magnitude of $\tau_\mathrm{m}^\mathrm{AR}$ and $\widetilde \tau_\mathrm{m}^\mathrm{\,AR}$ is much smaller than that of the measurements. These results suggest that the true travel-time perturbations caused by the inflows do not fully explain the solar-cycle variations in the travel-time measurements.

\section{Summary and discussion \label{sec:discussion}}

We used LCT flow maps over nine years during Cycle 24. We removed large-scale background flows and extracted the flows associated with active regions.
We averaged these flows in longitude to study their latitudinal profile and time evolution. 
The latitudinal flow exhibits an inflow pattern converging toward the active latitudes with a maximal peak-to-peak amplitude of $10$~m/s (yearly average) during solar maximum.
This amplitude is consistent with that found by \citet{2003Gizon}, \citet{2008GonzalezHernandez} and \citet{2020Komm}.
The longitudinal flow exhibits a pattern around the activity belts resembling that of the torsional oscillations. This pattern may be consistent with that observed by \citet{2020Komm}.
The maximum peak-to-peak amplitude is about $10$~m/s.
The amplitude and the structure of both flow components are correlated with the strength and the distribution of the magnetic activity.

We computed the corresponding forward helioseismic travel-time perturbations using 3D sensitivity kernels and using an arc-to-arc geometry in the north-south direction.
We assumed constant inflows with depth to place an upper limit on the contribution of inflows to helioseismic measurements of the meridional circulation. 
For separation distances $\Delta = 6^\circ$\,--\,$18^\circ$,
the extrema of the averaged forward travel-time perturbations are $\pm 0.3$~s at mid-latitudes during the peaks of solar activity.
We also averaged the forward travel-time perturbations with the masking of magnetic regions as in \citet{2020Gizon}, and found that the masking significantly reduced the amplitude of the travel-time perturbations, leading to extrema of about $\pm 0.1$~s during solar maximum.

We extended the active-region flows and the forward travel-time perturbations to cover two solar cycles, from May 1996 to April 2019, using a model based on the latitudinal gradient of the magnetic field strength. We assumed that the travel-time perturbations are roughly proportional to the flows. This simple model reproduces the inflow patterns throughout the solar cycle and allows us to compare our results with the measurements of \citet{2020Gizon} over two solar cycles. We found that the near-surface active-region flows do not explain in full the solar-cycle variations seen in the measurements of the meridional circulation.

We note that the background flows, defined as the flows that are far from active regions, are expected to represent the systematics, but they might also contain global-scale flows in the quiet regions, if any. We see that they exhibit a large-scale modulation throughout the solar cycle that seems correlated with the magnetic activity (Appendix \ref{sec:bkg_flows}).
Since the LCT data were filtered to remove only the periods of 24~hr, 1~yr, and the mean of the time series \citep{2017Loeptien,2021Gottschling}, long-term variations of the global-scale meridional flow could still remain in the background flows.
However, we cannot exclude that there exist systematics that vary with the solar cycle. Separating the true global-scale flows from the systematics in the LCT data is beyond the scope of this paper.
 
The fact that the surface active-region flows do not fully account for the observed solar-cycle variations of the meridional flow implies that there may be time-varying flows far from activity.
Other similar studies in which the active-region flows and quiet-region flows are separated seemed to confirm this.
\citet{2008GonzalezHernandez} used ring-diagram analysis to infer the subsurface meridional flow, and found that the inflows persist even after the flows surrounding the active regions are excluded. They attributed it to the fact that their masking may not remove the weaker magnetic regions and the diffuse field, but they also did not exclude the possibility that inflows may exist in quiet regions.
Similarly, \citet{2020Komm} studied the time variations of subsurface flows for quiet and active regions, separately, over the past two solar cycles and found a solar-cycle modulation in the quiet-Sun flows.

\citet{2010Hathaway} found that the meridional flow is stronger during solar minimum, although they did not disentangle the active-region flows from the global circulation. \citet{2018Lin} assumed that the meridional flows are simply a linear combination of the local inflows and global-scale meridional flows and found a similar result.
A number of other studies, in which the authors subtracted a time-averaged meridional flow profile, observed residuals during solar minima \citep[e.g.,][]{2011Hathaway,2015Komm,2021Getling}; however, the patterns of these residuals differ from one another, depending on the time periods used to compute the reference.
To avoid this dependence, \citet{2010GonzalezHernandez} subtracted a low-order polynomial fit and also found meridional flow residuals during solar minimum.
This might explain why the active-region flows only account for a fraction of the solar-cycle variations of the meridional flow. 

We note finally that other phenomena related to the surface magnetic activity can contribute to the temporal variation of the travel-time perturbations measured by \citet{2020Gizon}. In particular, the Woodard effect \citep{1997Woodard} can add systematics via the localized absorption of acoustic waves by sunspots. This effect has not been taken into account in our study.

\begin{acknowledgements}
P.-L. P. is part of the International Max Planck Research School. D.~F. and L.~G. acknowledge funding from the ERC Synergy Grant WHOLE~SUN \#810218. We thank B. Löptien for kindly accepting to provide the LCT data series, and N. Gottschling for fruitful discussions. The HMI data are courtesy of NASA/SDO and the HMI Science Team. SOHO is a project of international cooperation between ESA and NASA. The computational resources were provided by the German Data Center for SDO through grant 50OL1701 from the German Aerospace Center (DLR). The sunspot numbers are from WDC-SILSO, Royal Observatory of Belgium, Brussels.
\end{acknowledgements}

\bibliographystyle{aa}
\bibliography{my_bibliography}

\begin{appendix}

\section{Contour determination \label{sec:contour}}

\begin{figure}[h]
    \centering
    \includegraphics[width=0.5\textwidth]{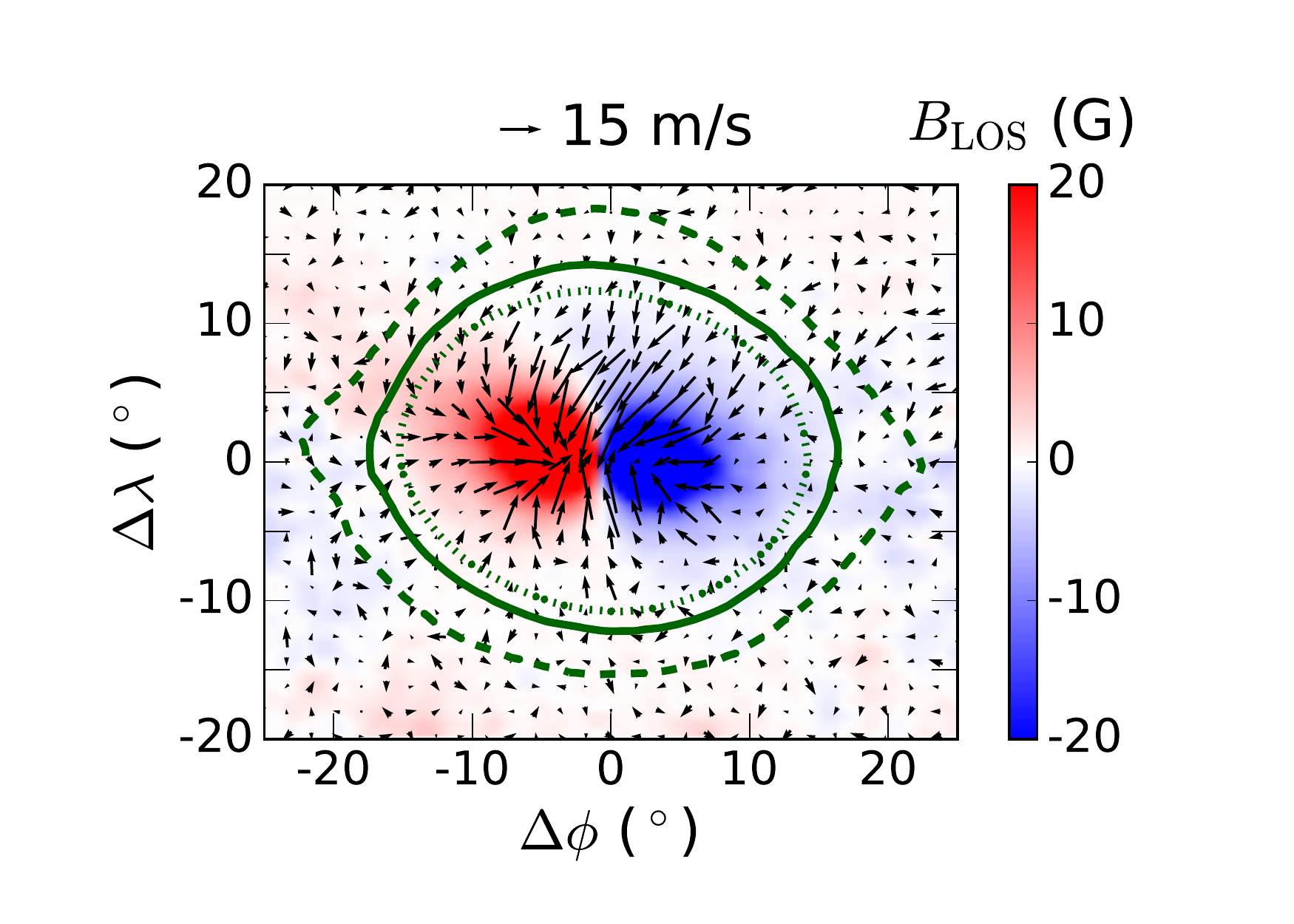}
    \caption{Ensemble-averaged active-region flows. $\Delta \phi$ and $\Delta \lambda$ are the longitude and latitude with respect to the center of the active region. The flows close to the center of each polarity have been masked out before averaging because LCT flows in highly-magnetic regions are not reliable \citep{2017Loeptien}. The background shows the ensemble-averaged line-of-sight magnetic field. 
    For a given value of $b$ that determines a contour for an individual active region, we draw the line delimiting the areas included in the contours for at least $90\%$ of the active regions used in the ensemble average; we try the following values of $b$: $4$~G, $3.5$~G, and $3$~G (dotted, solid, and dashed green lines, respectively).
    }
    \label{fig:figa1}
\end{figure}

We first compute the ensemble average of active regions in a way similar to \citet{2019Braun} and \citet{2021Gottschling}. A brief description of the procedure is as follows.
We track and remap the $B_r$ maps (obtained in Section~\ref{sec:contour_def}) in the same way as the flow maps in Section~\ref{sec:ar_flows}.
We compute the daily averages of unsigned magnetic flux density maps from $B_r$.
We smooth these maps by a 2D Gaussian with a FWHM of $10^\circ$ (heliographic degrees).
We identify the peaks (pixels of value higher than that of any of the eight neighboring pixels), and discard the ones that are within $20\degr$ of other peaks with stronger flux density.
This procedure ensures that all the selected peaks are clearly separated from each other.
We compute the total unsigned magnetic flux over a box, spanning $20\degr$ in longitude and $10\degr$ in latitude around each peak.
Only if the total unsigned magnetic flux is above $10^{21}$~Mx do we retain the active regions associated with the peaks.
We compute the center of mass of the pixels with positive $B_r$ and that of the pixels with negative $B_r$, both weighted by the smoothed and daily-averaged unsigned flux density; the average of the two centers of mass is defined as the center of an active region.
To limit the impact of noise, we do not consider pixels with $|B_r|$ less than $50$~G in the computation of the center of mass and of the magnetic flux.
We align the selected active regions with respect to their center.
If an active region is in the southern hemisphere, we flip it in the north-south direction and we reverse the sign of $B_r$ and $v^\text{AR}_\theta$,
so that it can be averaged together with the active regions in the northern hemisphere.

Figure~\ref{fig:figa1} shows the ensemble-averaged active-region flows using the $v^\text{AR}_\theta$ and $v^\text{AR}_\phi$ from Section~\ref{sec:ar_flows}. 
The inflow pattern is clearly seen, with the convergence center located preferentially in the trailing polarity. The pattern extends on average up to $10^\circ$ away from the center of each polarity, in both the latitudinal and the longitudinal directions. This is consistent with the results of, e.g., \citet{2017Loeptien}, \citet{2019Braun} and \citet{2021Gottschling}.

For each active region, we define the contour as the line along which the smoothed $|B_r|$ (see Section~\ref{sec:contour_def}) is equal to a magnetic threshold $b$.
Three values of $b$, $3.5$~G, $4$~G, and $4.5$~G, are tested.
For each value, we draw in Fig.~\ref{fig:figa1} the line delimiting the area included in the contours for at least $90\%$ of the active regions.
All three areas cover up to at least $10\degr$ away from the active regions, that is, they include most of the active-region inflows.

We want to choose the contour that extends as far away as possible from the active regions, to make sure that all the active-region flows are included.
However, with $b=3$~G, the background flows contain no data for more than a year at mid-latitudes around 2011 in the north and around 2014 in the south; in that case, smoothing the background flow maps in time with a Gaussian of FWHM of one year is not possible. Increasing the FWHM might also not be adequate because the time scale of variation of the background flows is on the order of years (see Appendix~\ref{sec:bkg_flows}). On the other hand, 
using $b=4$~G results in weaker inflows,
because the background flows may contain part of the outer edge of the inflows.
As a result, we choose $b=3.5$~G as the best compromise to consider as much active-region flows as possible and still have a reliable background estimation.

\section{Temporal variation of the background flows \label{sec:bkg_flows}}

\begin{figure*}[t]
    \centering
    \includegraphics[width=\textwidth]{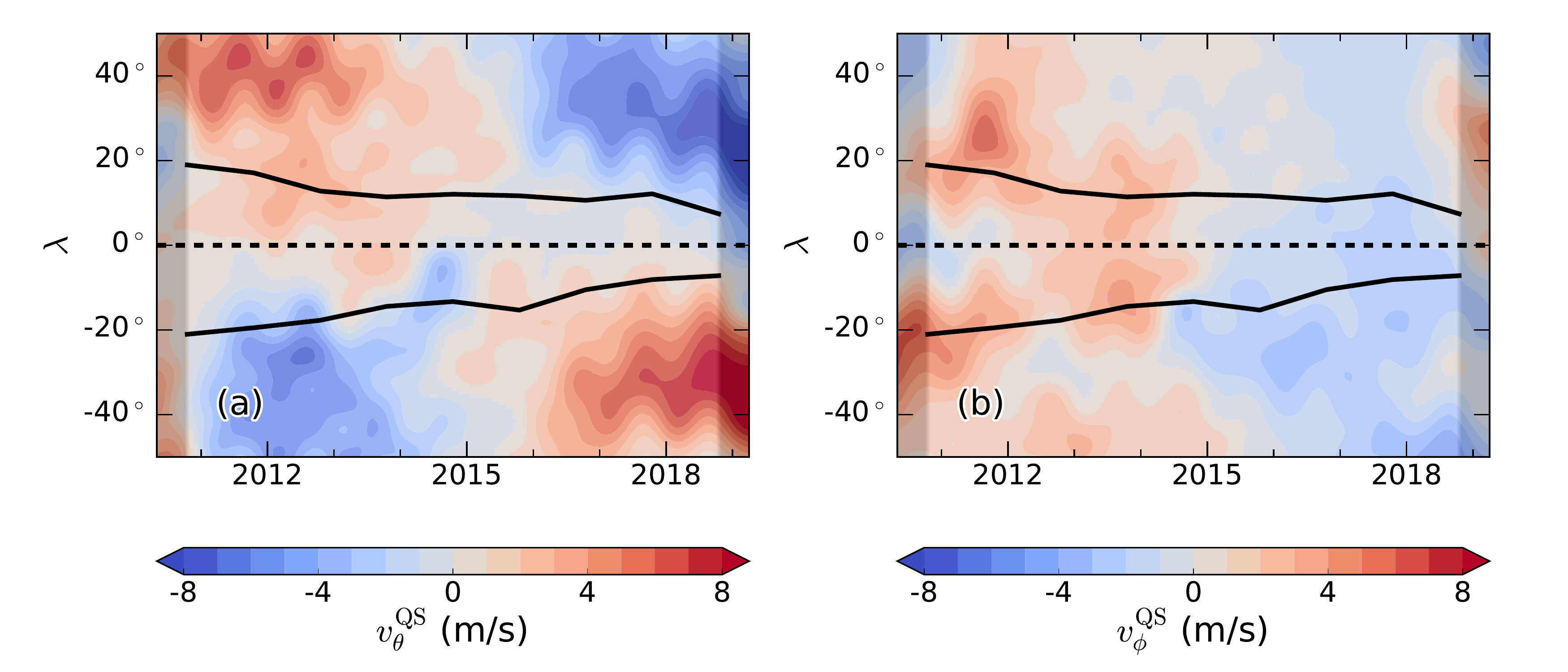}
    \includegraphics[width=\textwidth]{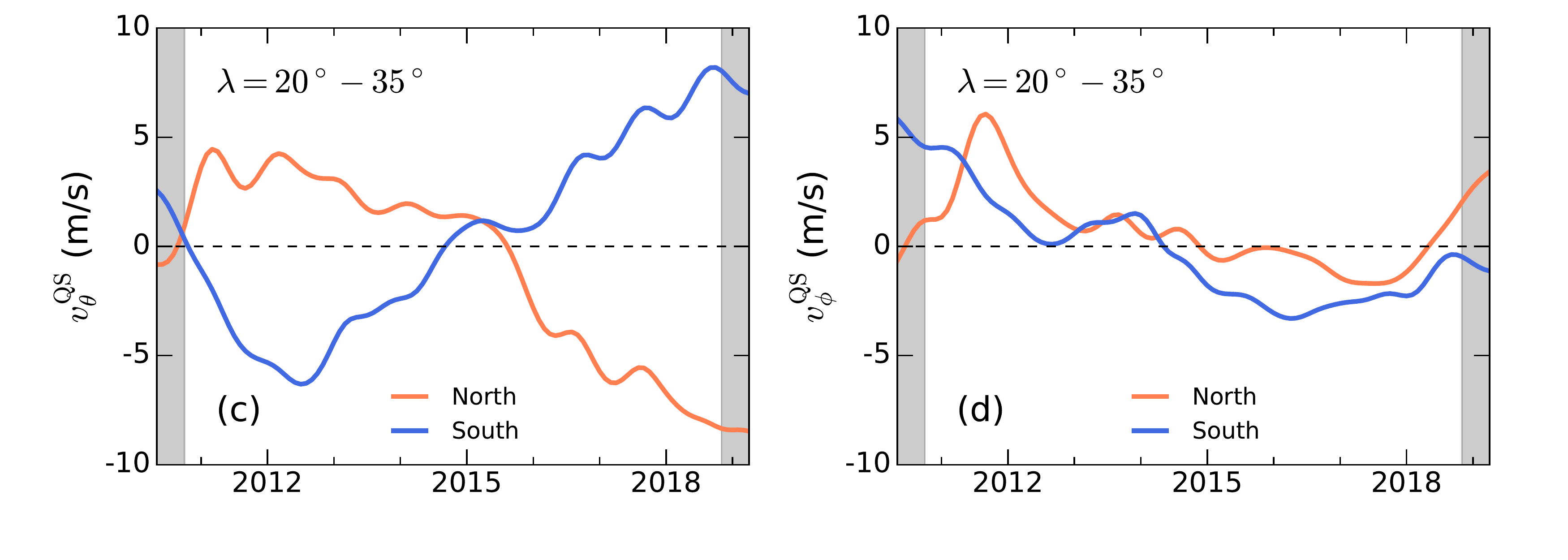}
    \caption{Longitudinally-averaged background flows $v_{\theta}^{\mathrm{QS}}$ (\textit{panel a}; positive values are southward) and $v_{\phi}^{\mathrm{QS}}$ (\textit{panel b}; positive values are prograde). We smoothed in latitude with a Gaussian of FWHM of $3.6^\circ$ and in time with a Gaussian of FWHM of $6$ months. \textit{Panels c--d}: Averages of the top panels over the latitudes $\lambda=20^\circ-35^\circ$ in each hemisphere. The typical standard errors of the mean, computed over the latitudinal interval, are about $0.06$~m/s for $v_{\theta}^{\mathrm{QS}}$ and about $0.03$~m/s for $v_{\phi}^{\mathrm{QS}}$, and are not shown here. In all panels, the shaded areas indicate the times when the edge effects become visible due to the one-year temporal smoothing of the flows done in Section~\ref{sec:bkg_smooth}.}
    \label{fig:figb1}
\end{figure*}

We track, remap and average the background flows in the same way as the active-region flows in order to study their temporal variation. The results are presented in Fig.~\ref{fig:figb1}.
A large-scale modulation of the meridional component is present, with a peak-to-peak amplitude of about $12$~m/s after we average over mid-latitudes (Fig.~\ref{fig:figb1}c). This trend is roughly antisymmetric with respect to the equator, and resembles an equatorward flow in the first half of the time series, then a poleward flow in the second half toward solar minimum (Fig.~\ref{fig:figb1}a). This is consistent with \citet{2021Gottschling}.
These flows contain systematics, but they might also contain true flows. If this is the case, this trend may indicate that the global-scale meridional flows become stronger when approaching the solar minimum, as was previously reported by, e.g., \citet{2010Hathaway}, \citet{2010GonzalezHernandez}, \citet{2015Komm}, \citet{2018Lin}, and \citet{2021Getling}.
 
The toroidal component also shows a modulation that is roughly symmetric with respect to the equator (Fig.~\ref{fig:figb1}b).
There exists a decreasing trend with time, with an amplitude that varies between $-3$ and $+5$~m/s (Fig.~\ref{fig:figb1}d).
We remind the reader that the mean value over the whole time period had been removed from the LCT data (see Section~\ref{sec:lct}), so the variation is with respect to the time average. 
The prograde pattern during the first half of the time series migrates toward the equator along with the activity belts, which resembles the behavior of torsional oscillations \citep[e.g.,][]{2020Komm}. This suggests that this component of the background flows might also contain
true flows in the quiet Sun.

We note that, as mentioned in Section~\ref{sec:bkg_smooth} and Appendix~\ref{sec:contour}, the background flow maps can contain little to no data at mid-latitudes
during active times. These cut-outs of active regions in the background flow maps are filled in using neighboring data with Gaussian smoothing. This smoothing procedure may result in a slight underestimate of the background flows at mid-latitudes.

\section{3D Born sensitivity kernels \label{sec:born}}

The computation of the 3D Born sensitivity kernels is based on the approaches from \citet{2017Gizon} and \citet{2018Fournier}. The wave field $\psi(\br,\omega)$ is solution of a scalar wave equation in the frequency domain
\begin{equation}
 -(\omega^2 + 2\ii \omega \gamma ) \psi - 2\ii\omega \bu \cdot \nabla \psi - c \nabla \cdot \left( \frac{1}{\rho} \nabla (\rho c \psi) \right) = s(\br,\omega), \label{eq:wave}
\end{equation}
where $\rho$ is the density, $c$ the sound speed, $\gamma$ the attenuation and $s$ a stochastic source term describing wave excitation. The wave field is related to the 3D wave displacement $\bxi$ through $\psi(\br,\omega) = c(\br) \nabla \cdot \bxi(\br,\omega)$. 

The flow kernels $\mathbf{K}=(K_r,K_\theta,K_\phi)$ can be computed from the knowledge of the Green's function solution of Eq.~\eqref{eq:wave} with a Dirac on the right hand side:
\begin{align}
 \mathbf{K}(\br; \br_1, \br_2) = 2 \ii \rho(r) \int_{-\infty}^{\infty} \omega & W^\ast(\br_1, \br_2, \omega) \Bigl[ G(\br_2;\br,\omega) \nabla C(\br_1;\br,\omega) \nonumber \\
 &- G^\ast(\br_1;\br,\omega) \nabla C^\ast(\br_2;\br,\omega) \Bigr] \;\textrm{d}\omega, \label{eq:kernel}
\end{align}
where $\br_1=(R,\theta_1,\phi_1)$ and $\br_2=(R,\theta_2,\phi_2)$ are two observation points, $R$ is the observation radius, $C$ is the cross-covariance, $W$ is a weighting function in order to relate the travel-time perturbation to changes in the cross-covariance, and $\nabla$ is the gradient operator with respect to $\br$. As in \citet{2017Gizon}, we assume energy equipartition so that the cross-covariance is related to the imaginary part of the Green's function. For the sake of simplicity, we now drop the $\omega$ in the notation of the cross-covariance and of the Green's function. In a spherically symmetric background, the Green's function depends only on the angular distance between source and receiver and can be obtained from its Legendre coefficients \citep{2018Fournier}:
\begin{equation}
 G(r,\theta,\phi; \br_i) = \frac{1}{\sqrt{2\pi}} \sum_{\ell}  G_\ell(r; R) P_\ell(\cos\gamma_i), \label{eq:Glm}
\end{equation}
where $i \in \{1,2\}$, $P_\ell$ is the Legendre polynomial of order $\ell$, 
\begin{equation}
 \cos\gamma_i = \cos\theta \cos\theta_i + \sin\theta \sin\theta_i \cos(\phi-\phi_i), \label{eq:cosgamma}
\end{equation}
and $G_\ell(r;R)$ is solution of
\begin{equation}
    -(\omega^2 + 2\ii \omega \gamma ) G_\ell - c \frac{d}{dr} \left( \frac{1}{\rho} \frac{d}{dr} (\rho c G_\ell) \right) + \frac{\ell(\ell+1) c^2}{r^2} G_\ell = \delta(r-R).
\end{equation}
We keep the values of $\ell$ up to $300$.
 
Inserting Eq.~\eqref{eq:Glm} into Eq.~\eqref{eq:kernel}, the kernel for the radial flow is given by
\begin{align}
 K_r(\br; \br_1, \br_2) = \frac{2 \rho(r)}{\pi} \sum_{\ell, \ell'} \Bigl[ & - f_{\ell \ell'}^r(r) P_{\ell'}(\cos\gamma_1) P_{\ell}(\cos\gamma_2) \nonumber \\
 & + g_{\ell \ell'}^r(r)  P_{\ell}(\cos\gamma_1) P_{\ell'}(\cos\gamma_2) \Bigr], \label{eq:Kr3D}
\end{align}
where
\begin{align}
 f_{\ell \ell'}^{r}(r) &=  \int_{0}^{\infty} \omega \textrm{Im} \left[ W^\ast(\br_1, \br_2, \omega) G_\ell(r; R) \partial_r C_{\ell'}(r; R) \right] \dd \omega, \\
 g_{\ell \ell'}^{r}(r) &=  \int_{0}^{\infty} \omega \textrm{Im} \left[ W^\ast(\br_1, \br_2, \omega) G_\ell^\ast(r; R) \partial_r C_{\ell'}^\ast(r; R) \right] \dd \omega.
\end{align}

Similarly, denoting
\begin{align}
 f_{\ell \ell'}^{j}(r) &= \int_{0}^{\infty} \omega \textrm{Im} \left[ W^\ast(\br_1, \br_2, \omega) G_\ell(r; R) C_{\ell'}(r; R) \right] \dd \omega, \\
 g_{\ell \ell'}^{j}(r) &=  \int_{0}^{\infty} \omega \textrm{Im} \left[ W^\ast(\br_1, \br_2, \omega) G_\ell^\ast(r; R)  C_{\ell'}^\ast(r; R) \right] \dd \omega,
\end{align}
where $j \in \{\theta,\phi\}$, the kernels for the horizontal flow components are given by
\begin{align}
 K_{j}(\br; \br_1, \br_2) = \frac{2 \rho(r)}{\pi \; r} \sum_{\ell, \ell'} \Bigl[ & -f_{\ell \ell '}^j(r) \alpha_1^{j} P'_{\ell'}(\cos\gamma_1) P_{\ell}(\cos\gamma_2) \nonumber \\
 & + g_{\ell \ell'}^j(r)  \alpha_2^{j} P_{\ell}(\cos\gamma_1) P'_{\ell'}(\cos\gamma_2) \Bigr],  \label{eq:Kth3D}
\end{align}
where $P'$ is the derivative of the Legendre polynomials and
\begin{align}
 \alpha_i^\theta &= -\cos\theta_i \sin\theta - \sin\theta_i \cos\theta \cos(\phi-\phi_i), \\
 \alpha_i^\phi &= -\sin\theta_i \sin(\phi-\phi_i)
\end{align}
are respectively the derivative of $\cos\gamma_i$ with respect to $\theta$, and the derivative with respect to $\phi$ and divided by $\sin\theta$.

Let's consider $\br_1$ and $\br_2$ two foot points in the arc-to-arc geometry described by \citet{2017Liang} and used in this paper. Let's denote $(\theta_0,\phi_0)$ the colatitude and longitude of the midpoint, $\Delta$ the separation distance between the foot points, and $\psi$ between a meridian and the ray path connecting the paired points on the arcs. Then we rewrite the flow kernels as
\begin{equation}
    \mathbf{K}(\br; \br_1, \br_2) = \mathbf{K}(\br; \theta_0,\phi_0,\Delta,\psi).
\end{equation}
We use this notation in the main text in order to make the averaging of the travel-time perturbations over the arcs more explicit.

\section{Flow and travel-time modeling \label{sec:modeling}}
\begin{figure}[h]
    \centering
    \includegraphics[width=0.5\textwidth]{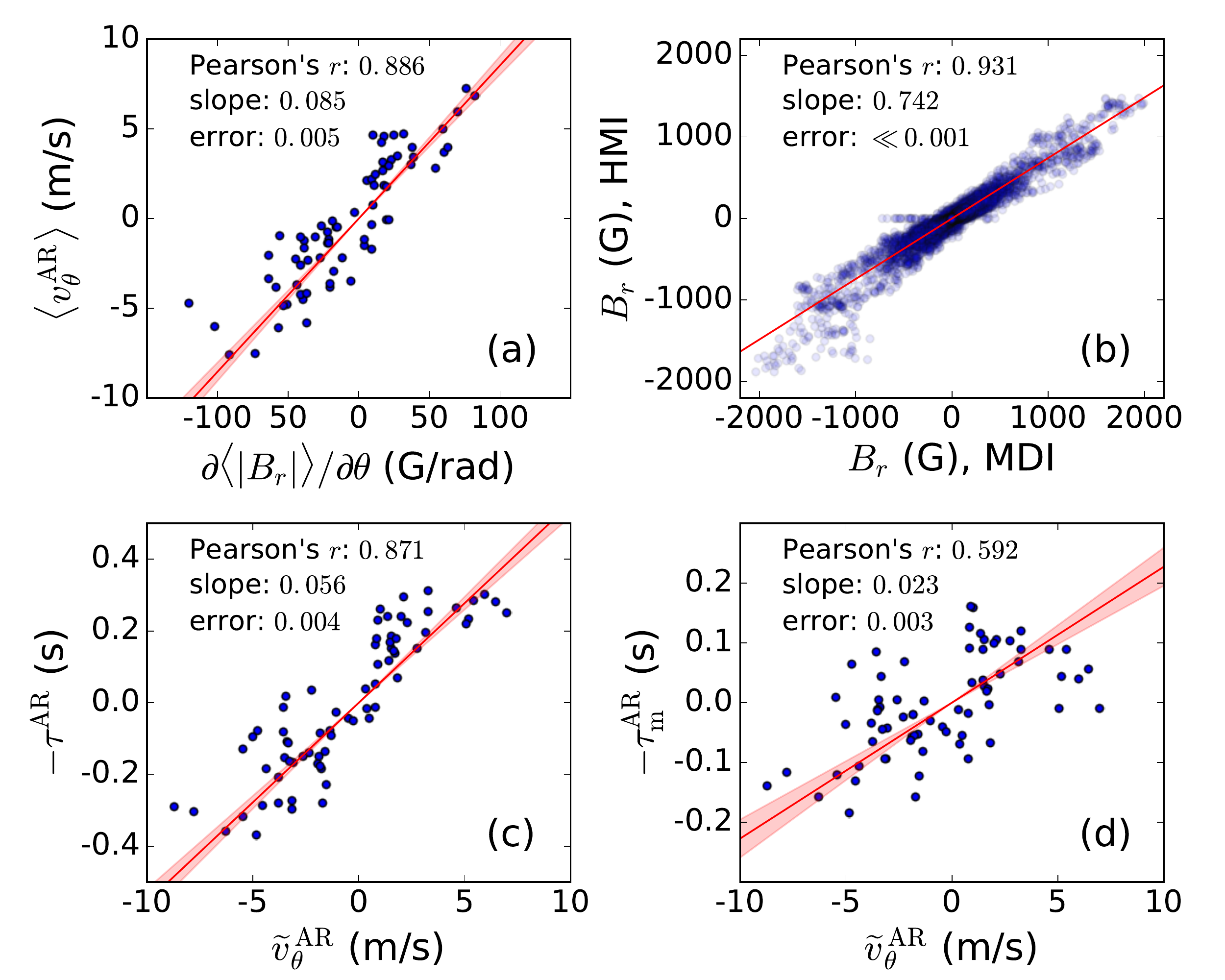}
    \caption{Scatter plots of the data used for the determination of the proportionality constant $c_0$ (\textit{panel a}), of the scaling factor for the magnetograms (\textit{panel b}), of the conversion constant for $\widetilde \tau^\mathrm{\,AR}$ (\textit{panel c}), and the conversion constant for $\widetilde \tau_\mathrm{m}^\mathrm{\,AR}$ (\textit{panel d}). The red lines indicate the best fits. The red shaded areas represent the errors on the slopes, obtained from the fits. The numbers in the top left corners are the values of the Pearson correlation coefficient, of the slope, and of the error on the slope. In all cases, the $p$ values of the correlation coefficients are too small to be given here ($\ll 0.01$).}
    \label{fig:figd1}
\end{figure}

The linear fitting procedure in Section~\ref{sec:extension} takes into account the errors in both coordinates \citep[][Section~15.3]{2007Press}.
We present the results of the fits in Fig.~\ref{fig:figd1}. For Fig.~\ref{fig:figd1}a,c,d, because the data used were smoothed in time and latitude, we use the points distant from each other by $6$~months and $3.6\degr$ in latitude so that they are independent. In all panels, the value of the intercept is smaller than the error in the vertical coordinate estimated from the misfit, so we neglected it.

There is a clear linear correlation between $\partial \langle |B_r| \rangle/\partial \theta$ and $\langle v_\theta^\mathrm{AR} \rangle$ (Fig.~\ref{fig:figd1}a), between the HMI magnetic field and the MDI magnetic field (Fig.~\ref{fig:figd1}b), and between $\widetilde v_\theta^\mathrm{\,AR}$ and $\tau^\mathrm{AR}$ (Fig.~\ref{fig:figd1}c). In contrast, the correlation between $\widetilde v_\theta^\mathrm{\,AR}$ and $\tau_\mathrm{m}^\mathrm{AR}$ (Fig.~\ref{fig:figd1}d) is not as high in other panels, implying that a linear model might yield a poor fit. This is because we cannot mask magnetic pixels in the modeled flows, which are a function of time and latitude but not longitude. Therefore we have to use the same flows for the fit of $\tau_\mathrm{m}^\mathrm{AR}$ as for the fit of $\tau^\mathrm{AR}$. The larger relative error in the slope (shaded area in red) in Fig.~\ref{fig:figd1}d reflects the poor linear fit.

\end{appendix}

\end{document}